\documentclass[conference]{IEEEtran}

\usepackage{calrsfs}
\DeclareMathAlphabet{\pazocal}{OMS}{zplm}{m}{n}
\usepackage{epsfig,endnotes,xspace,hyperref,url,diagbox}
\AtBeginDocument{}
\usepackage{amsmath,amsthm,amssymb}
\usepackage{tikz}
\usepackage{enumerate}
\usepackage{varioref}
\usepackage{graphicx}
\usepackage{epstopdf}
\usepackage{color}
\usepackage{xcolor}
\usepackage{colortbl}
\usepackage{multirow}
\usepackage{comment}
\usepackage[utf8]{inputenc}
\usepackage{mathtools}
\usepackage{wrapfig}
\usepackage{booktabs}
\usepackage{caption}
\usepackage{rotating} 
\usepackage{enumitem}
\usepackage{subcaption}
\usepackage{float}

\newcommand{\todo}[1]{{\color{red} \textbf{Todo:}\{ \emph{#1} \}}}
\newcommand{\fixme}[1]{{\color{red} \textbf{Fixme:}\{ \emph{#1} \}}}
\newcommand{\says}[2]{{\color{blue}{#1 says: }{#2}}\xspace}
\newcommand{\ap}[1]{\says{aiping}{#1}}
\newcommand{\tw}[1]{\says{tianhao}{#1}}

\newcommand{\mypara}[1]{\vspace*{0.05in}\noindent\textbf{#1}.$\;$}
\newcommand{\question}[1]{\vspace*{0.05in}\noindent\textbf{#1}.$\;$}

\newtheorem{definition}{Definition}

\newcommand{\Domain}{\ensuremath{D}\xspace}

\renewcommand{\Pr}[1]{\ensuremath{\mathsf{Pr}\left[#1\right]}\xspace}

\renewcommand{\AA}{\mathbf{A}}

\renewcommand{\Pr}[1]{\ensuremath{\mathsf{Pr} \left[#1\right] }\xspace}

\newcommand{\tuple}[1]{\ensuremath{\langle #1\rangle}\xspace}

\newcommand{\mytitle}{{\huge Using Illustrations to Communicate \\ Differential Privacy Trust Models}\\ {\Large An Investigation of Users' Comprehension, Perception, and Data Sharing Decision}}

\newcommand{\results}{\ensuremath{\mathsf{\mathbf{R}}}\xspace}

\title{\mytitle}

\begin{document}
\pagestyle{plain}

\author{

	{\rm Aiping Xiong\textsuperscript{1}}
	\and 
	{\rm Chuhao Wu\textsuperscript{1}}
	\and 
    {\rm Tianhao Wang\textsuperscript{2}}
	\and
	{\rm Robert W. Proctor\textsuperscript{3}}
	\and
	{\rm Jeremiah Blocki\textsuperscript{3}}
	\and
	{\rm Ninghui Li\textsuperscript{3}}
	\and
	{\rm Somesh Jha\textsuperscript{4}}
	\and
	\;\;\textsuperscript{1}\textit{Pennsylvania State University}
	\and
    \textsuperscript{2}\textit{University of Virginia}
    \and
    \textsuperscript{3}\textit{Purdue University}
    \and
    \textsuperscript{4}\textit{University of Wisconsin-Madison}
    
}

\maketitle

\begin{abstract}

Proper communication is key to the adoption and implementation of differential privacy (DP). However, a prior study found that laypeople did not understand the data-perturbation processes of DP and how DP noise protects their sensitive personal information.  Consequently, they distrusted the techniques and chose to opt out of participating.

In this project, we designed explanative illustrations of three DP models (Central DP, Local DP, Shuffler DP) to help laypeople conceptualize how random noise is added to protect individuals' privacy and preserve group utility.  Following pilot surveys and interview studies, we conducted two online experiments ($N=595$) examining participants' comprehension, privacy and utility perception, and data-sharing decisions across the three DP models. Besides the comparisons across the three models, we varied the noise levels of each model. 
We found that the illustrations can be effective in communicating DP to the participants.  Given an adequate comprehension of DP, participants preferred strong privacy protection for a certain type of data usage scenarios (i.e., commercial interests) at both the model level and the noise level. 
We also obtained empirical evidence showing participants' acceptance of the Shuffler DP model for data privacy protection. 
Our findings have implications for multiple stakeholders for user-centered deployments of differential privacy, including app developers, DP model developers, data curators, and online users.  

\end{abstract}

\section{Introduction}
\label{sec:intro}


\textit{Differential Privacy} (DP, also called \textit{Central DP})~\cite{Dwo06} is a promising approach to preserve privacy with a quantifiable protection guarantee and acceptable utility in the context of statistical information disclosure.  Specifically, it adds noise to the aggregated-level results such that an individual's information disclosure is bounded.  The US Census Bureau has implemented Central DP to protect the privacy of each participant of the 2020 Census~\cite{uscensus}. 


In recent years, local differential privacy (\textit{Local DP})~\cite{duchi2013local,evfimievski2004privacy,kasiviswanathan2011can,Warner65} has become popular because of its deployment in companies such as Google~\cite{rappor}, Apple~\cite{apple-dp}, and Microsoft~\cite{nips:DingKY17}. Local DP differs from Central DP in that random noise is added 
by each user before sending the data to the server.  
Thus, users do not need to rely on a trusted third party.  
Nevertheless, removing the trusted central party comes at the cost of utility.  Since every user adds some independently generated noise, the effect of noise adds up when aggregating the result.  As a result, while noise of scale (standard deviation) $\Theta(1)$ suffices for Central DP, Local DP has noise of scale $\Theta(\sqrt{n})$~\cite{chan2012optimal} on the aggregated result ($n$ is the number of users). 

More recently, researchers introduced {\it Shuffler DP}~\cite{balle2019privacy,cheu2018distributed,erlingsson2019amplification}, which achieves a middle ground between Central DP and Local DP.
Shuffler DP involves an auxiliary party called the shuffler. Users send their perturbed data to the shuffler; the shuffler shuffles the users' data, and then send data to the server, and thus removing the linkage between users and their reports.  Because of this anonymity property, users can add less noise while achieving the same level of privacy.  
The downside of the shuffler DP is that it requires that the shuffler should not collude with the server (otherwise, the user obtains privacy protection only corresponding to the Local DP noise, and there is no benefit of shuffling). Google has deployed a shuffler DP model Prochlo~\cite{bittau2017prochlo}.

With the increasing deployment of DP and its variants, 
research has been conducted to examine whether users can understand these techniques, trust them, and consequently, increase their willingness to share data when the deployment of those techniques is communicated~\cite{bullek2017towards,xiong2019arxiv,cummings2021need}. 
Using {\it textual descriptions}, Xiong et al.~\cite{xiong2019arxiv} conducted a series of online human-subject experiments with Central DP and Local DP in a health-app data collection setting.  Among various descriptions, they found that the descriptions of implications helped laypeople understand that Local DP provides better privacy protection than Central DP, and facilitated laypeople's information disclosure decisions.  The results also revealed that participants had difficulty understanding the data perturbation processes, especially how random noise protects personal sensitive information.  Yet, a major shortcoming in Xiong et al.~\cite{xiong2019arxiv} (and others~\cite{bullek2017towards,cummings2021need}) is that they mainly focus on communicating privacy protection of DP, but ignore the dimension of utility (i.e., a reduction of the usefulness or accuracy of data), which comes with the privacy protection.  

\begin{figure*}[t]
\centering
\begin{subfigure}[b]{.55\linewidth}
    \centering
    \includegraphics[width=\linewidth]{fig/Shuffler_image.png}
    \vspace{-0.1cm}
    \caption{Data flow diagram of Shuffler DP.}
    \label{fig:diagram_ShufflerDP}
\end{subfigure}
\begin{subfigure}[b]{0.45\linewidth}
    \centering
    \includegraphics[width=\linewidth]{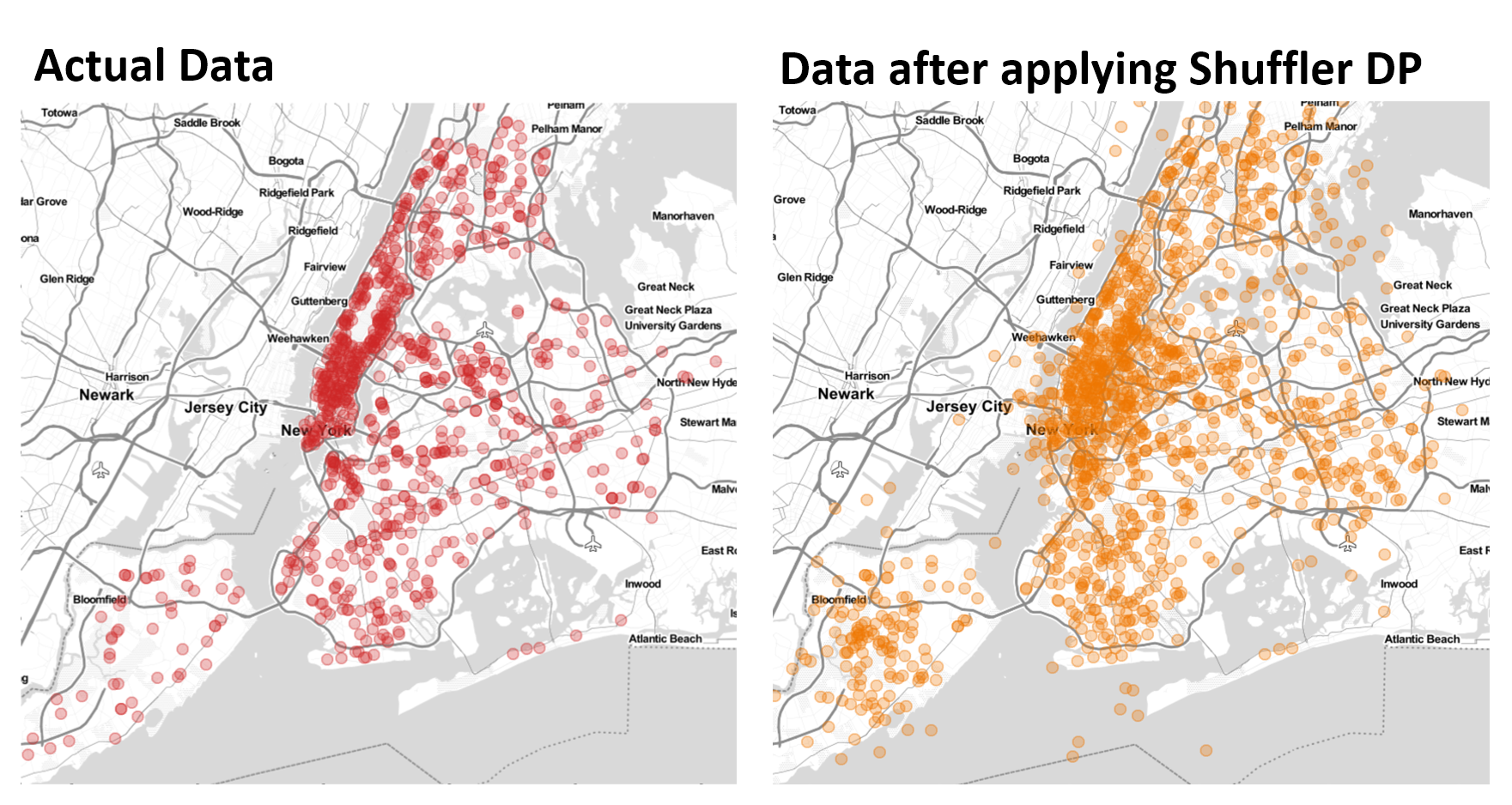}
    \caption{Illustration of utility cost for the Shuffler DP.}
    \label{fig:data_ShufflerDP}
\end{subfigure}
\vspace{-0.1cm}
\caption{Illustration of the Shuffler DP in Experiment 1. 
{{
Fig.~\ref{fig:diagram_ShufflerDP} depicts the processes of data perturbation and shuffling using Shuffler DP. 
Fig.~\ref{fig:data_ShufflerDP} illustrates the utility cost for the Shuffler DP model, where darker dots indicate more individual users\protect\footnotemark.
}}}
\label{fig:diagram_ShufflerDP_all}
\vspace{-0.3cm}
\end{figure*}

The goal of this work is to understand how to effectively communicate DP to end users and consequently inform their online information disclosure. 
Specifically, we focus on the following three research questions (RQs): 

\begin{itemize}[leftmargin=*]
    
    \item \mypara{RQ1} 
    Given the comprehension of privacy and utility tradeoff of DP, how do participants' perceived utility cost and privacy protection vary across the three DP models?
    
    \item \mypara{RQ2} 
    Does data usage (i.e., public good or commercial interests) impact participants' data-sharing decisions? 
    
    \item \mypara{RQ3} 
    Do participants prefer different noise levels across the data usage scenarios and the three DP models?

    
    
    
    
    
\end{itemize}

\footnotetext{Data was generated based on the open dataset provided by the Department of Health and Mental Hygiene at https://data.cityofnewyork.us/Health/New-York-City-Locations-Providing-Seasonal-Flu-Vac/w9ei-idxz}
We propose to communicate the {\it privacy-utility tradeoff} of three DP models (Central DP, Local DP, and Shuffler DP) in the context of {\it location privacy setting}. To foster users' comprehension and consideration of the privacy-utility tradeoff across the three DP models, we developed an \textit{explanative illustration}~\cite{mayer1990illustration} for each model, in which verbal information in natural language and symbolic graphics were presented to illustrate DP privacy protection at an individual level and utility cost at an aggregated level.  Considering the abstract nature of the data perturbation process in DP models, especially the transformation of the data before and after the noise addition, we also designed \textit{stepwise} illustrations to facilitate participants' mental representation process and comprehension of the illustrations.  

Since DP is a concept with which the public is not familiar, we also proposed comprehension questions to ensure participants’ basic understanding of the DP models. We tried to exclude the impacts from extraneous factors by making the illustrations and surveys consistent among the three models. 
We constructed various scenarios to approximate real-world location data collection and use. 

We conducted two online surveys addressing the above {RQs}. 
To finalize the illustrations, we conducted a pilot survey and an interview study before each experiment. Results of those preliminary studies helped us propose the stepwise illustrations. 
In Experiment 1, we focused on comparing the three DP models using a within-subject design since users' comprehension of Shuffler DP has not been examined. 
We examined participants' perceptions of privacy protection and utility cost of the three DP models (\textbf{RQ1}). After viewing each data usage scenario, participants were prompted to select a preferred DP model for data disclosure (\textbf{RQ2}).
We conducted Experiment 2 to further investigate participants' privacy protection and utility cost of the three models using a between-subject design (\textbf{RQ1}). We also proposed illustrations of noise level and examined participants' preferred level of noise for data disclosure (\textbf{RQ3}). 
We obtained answers for each question as follows.

\begin{itemize}[leftmargin=*]

    \item \mypara{RQ1: Comprehension and Perception} 
    Participants in Experiment 1 showed a better understanding of the privacy or utility implications for the Central DP (73.7\%) and the Shuffler DP (76.6\%) than for the Local DP (49.2\%, see Sec~\ref{sec:formal_comp}). Yet, such differences across models were not significant in Experiment 2 (66.1\% on average, see Sec~\ref{sec:formal_comp_2}).  
    
    To guarantee the quality of the following perception and data-sharing responses, we only considered participants who at least answered half of the comprehension questions correctly.
    When participants had an adequate comprehension of privacy and utility implications, 
    the perceived utility showed no difference across the three models in both experiments (see Sec~\ref{sec:formal_perception} and Sec~\ref{sec:formal_percp_2} ). While the participants in Experiment 2 gave similar perceived privacy ratings across the three models, those in Experiment 1 gave the highest perceived privacy rating for the Local DP ($5.66$), followed by the Shuffler DP ($5.35$) and the Central DP ($4.56$). 
    
    \item \mypara{RQ2: Data-sharing Decisions} Participants in Experiment 1 showed more willingness to share data for the public-good usage (with an average rating score of $5.05$) than for the commercial-interests usage ($4.60$).  Regardless of data usage, participants showed more willingness of data disclosure with the Local DP ($5.06$) or the Shuffler DP ($4.96$) than with the Central DP ($4.44$), revealing their preference for stronger privacy protection (see Sec~\ref{sec:formal_decision}).
    \item \mypara{RQ3: Noise-level selection} In Experiment 2, we obtained the same effect of data usage: participants preferred high-level noise for the commercial-interests usage ($40.2\%$) than for the public-good usage ($29.1\%$, see Sec~\ref{sec:formal_noise_2}). In agreement with the results of Experiment 1, such preference was also model-independent. Yet, participants showed no preference differences in the noise level across the three models, consistent with the perceived privacy rating results. 
    
    
\end{itemize}

\vspace{0.1cm}To summarize, this work makes the following contributions to human-centered DP deployment. 

\begin{itemize}[leftmargin=*]

    \item We propose a novel method to communicate different DP trust models through illustrations, and examine participants’ comprehension, perceived utility/privacy, and data disclosure. 
    
    \item We identify that accurate perception of the privacy protection of different DP trust models and consequent informed decision are based on adequate comprehension of DP. 
    
    \item Regardless of the trust models or noise levels, we find that participants prefer stronger privacy protection for the commercial-related data usage scenarios than for the public-good scenarios. 
    
    \item We provide the first empirical evidence showing people's acceptance of the Shuffler DP model for data-privacy protection. 
    
    \item We obtain participants' preference of models with strong privacy protection in a within-subject design but not a between-subject design, indicating the importance of making different trust models available for users' informed privacy decisions. 
    
\end{itemize}

\section{Background and Related Work}
In this section, we discuss the three models of differential privacy, location privacy, prior works on DP communication, and privacy-utility tradeoff. 

\subsection{Three Models of Differential Privacy}
\label{subsec:dp}
Differential privacy~\cite{Dwo06} (DP) is a rigorous notion about individuals' privacy.  Intuitively, the DP notion requires a randomized algorithm that adds ``noise'' to the output, so that the impact of any single element in a dataset is ``obscured'' by noise. 

In what follows, we review the three models of DP.  

\subsubsection{Central Differential Privacy}
\label{subsec:cdp}
The classic DP notion works in the setting where there is a trusted data curator, who gathers data from individual users, processes the data in a way that satisfies DP, and then publishes the results.  To differentiate the classic notion from variants that are proposed later, we call it Central DP.

Denote a dataset as $D=\tuple{v_1, v_2, \ldots, v_n}$.  Two datasets $D=\tuple{v_1, v_2, \ldots, v_n}$ and $D'=\tuple{v'_1, v'_2, \ldots, v'_n}$ are said to be neighbors, or $D\simeq D'$, iff there exists at most one $i\in [n]=\{1,\ldots, n\}$ such that $v_i\neq v'_i$, and for other $j\neq i, v_j=v'_j$.  When $\delta=0$, we simplify the notation and call $(\epsilon, 0)$-Central DP as $\epsilon$-Central DP.  

\begin{definition}[Central Differential Privacy] \label{def:dp}
An algorithm $\AA$ satisfies $(\epsilon, \delta)$-Central DP, where $\epsilon, \delta \geq 0$,
if and only if for any neighboring datasets $D$ $\cong$ $D'$, and any set \results of possible outputs of $\AA$, we have
\begin{equation*}
\Pr{\AA(D)\in \results} \leq e^{\epsilon}\, \Pr{\AA(D') \in \results} + \delta
\end{equation*}
\end{definition}


\subsubsection{Local Differential Privacy}
\label{subsec:ldp}
Given the possibility of untrustworthy data curators, Local differential privacy (Local DP) has been proposed~\cite{duchi2013local,kasiviswanathan2011can}.  
Compared to the centralized setting, the local version of DP offers a stronger level of protection because noise is added on the user side before sending the data to a curator.  Since each user only reports the perturbed data, each user's privacy is still protected even if the aggregator is malicious. 

In the local setting, each user perturbs the input value $v$ using an algorithm $\AA_l$ and reports $\AA_l(v)$ to the aggregator.  

\begin{definition}[Local Differential Privacy] \label{def:ldp}
	An algorithm $\AA_l(\cdot)$ satisfies $(\epsilon, \delta)$-Local DP, where $\epsilon, \delta \geq 0$,
	if and only if for any input $v, v' \in \Domain$, and any set \results of possible outputs of $\AA$, we have
    \begin{equation*}
\Pr{\AA_l(v)\in \results} \leq e^{\epsilon}\, \Pr{\AA_l(v')\in \results} + \delta
	\end{equation*}
\end{definition}
Typically, the $\delta$ value used is $0$ (thus $\epsilon$-Local DP).  While Local DP provides a better privacy model (in that users do not need to send their sensitive data directly to the server), the total noise seen by the server is $\Theta(\sqrt{n})$~\cite{chan2012optimal} (which is much larger compared to the $\Theta(1)$ noise in Central DP), because each user must add 
noise independently of other users.


\subsubsection{Shuffling Differential Privacy}
The shuffling idea first appeared in Prochlo~\cite{bittau2017prochlo}, where a shuffler is inserted between the users and the server to break the linkage between the report and the user identification.  
The formal proof of the privacy benefit was given in~\cite{balle2019privacy,cheu2018distributed,erlingsson2019amplification}.  
In this model, each user adds Local DP noise to data, encrypts it, and then sends it to one new party called the shuffler.  The shuffler permutes the users' reported data, and then sends them to the server.  Finally, the server decrypts the reports and obtains the result.  
In this process, the shuffler only knows which report comes from which user, but does not know the content of the user's report.  On the other hand, the server cannot link a user to a report because the reports are shuffled.
The Shuffler DP model can be thought of as a model between Central DP and Local DP: Users' data sent to the server is protected, while the noise seen from the server is close to $\Theta(1)$.  However, it requires that the server and the shuffler do not collude with each other.  Note that while theoretically the shuffler can be composed of many servers and as long as one server is not colluding, the whole model is safe, in practice, introducing more servers also introduces more communication cost and other maintenance issues.  In this paper, we assume the shuffler is one server.

\subsection{Location Privacy}
Since the advances of sensor-based devices, such as wearable devices and smart phones, detailed user location data can be collected and examined to determine users' preference, as well as target them with services and advertisements. Although the information collected is not tied to any user's name or phone number but a unique ID (e.g., Google Chrome~\cite{uniqueid}), those with access to the raw data — including employees or third-party clients — could still identify one specific user without consent by using other related information~\cite{nylocation2018}. Moreover, the continual release of locations can be used as a trajectory, creating more emerging issues~\cite{chow2011trajectory}. 

Location privacy is a particular type of information privacy defined as the ability to prevent other parties from learning one's current or past location~\cite{beresford2003location}.  Much work of DP has been conducted in the location privacy context~\cite{andres2013geo,to2014framework}.  For example, Andr{\'e}s and his colleagues~\cite{ABCP13} proposed geo-indistinguishability, a differentially private location-based system to protect an individual's exact location, while maintaining the desired service with enough location data being disclosed.  The main idea is to a add controlled random noise to the radius \textit{r} that the individual has visited. For any radius larger than 0, an individual will have guaranteed privacy that depends on \textit{r}.

Researchers have designed interfaces to explain location privacy protection, such as LP-Guardian~\cite{fawaz2014location} and PrivacyGuard~\cite{song2015privacyguard}. They have also documented users' concerns~\cite{barkhuus2003location}, preferences~\cite{benisch2011capturing,lin2013comparative}, and behaviors~\cite{fisher2012short} relating to location privacy. To our knowledge, no work has
focused on explaining the DP techniques in the location setting nor evaluating how such communication impacts
users’ perception and data disclosure. We address this knowledge gap.  

\subsection{Differential Privacy Communication}
Bullek et al.~\cite{bullek2017towards} illustrated the randomized response technique~\cite{warner1965randomized} for Local DP using spinners and evaluated participants' preference of the privacy parameter in an online study. Each participant selected and experienced the perturbation for sensitive questions with three probabilities, corresponding to three $\epsilon$ values. Participants were asked to select a perturbation probability for a final high-sensitive question. Results of the online study with 228 participants showed that 75\% of them chose the largest perturbation, indicating a preference for strong privacy protection.
Prior studies using textual descriptions showed that the communication of differential privacy should focus on explaining how random noise protects individuals' information privacy~\cite{xiong2019arxiv}.  
A recent survey study examined the impact of six different textual descriptions of DP on participants' expectations for privacy and their willingness to share different kinds of information~\cite{cummings2021need}. 
Regardless of the descriptions, the results showed that informing participants of DP deployment did not raise their potential willingness to share the information. 



\subsubsection{Illustrations}
Previous studies found that learning from illustrated text produced better performance than learning from text alone in various educational settings~\cite{levie1982effects,morrison2001effectiveness}.  Dual coding theory~\cite{clark1991dual,paivio2006dual} also indicates that conveying information in both verbal and non-verbal (e.g., pictorial codes) representations provides double routes for the processing, encoding, and retrieval of the presented information.  
We developed an \textit{explanative illustration}~\cite{mayer1990illustration} for each model, in which verbal information in natural language and symbolic graphics was presented to promote the comprehension and consideration of the privacy-utility tradeoff across the three DP models.  The use of spaces in graphics for representing relevant elements and their relations also leveraged the power of spatial reasoning and inference in the human cognitive system~\cite{tversky2001spatial}. 

Moreover, with techniques, such as Central DP, a company can still collect raw data from individuals, indicating the compromise risk about which individuals were most concerned~\cite{xiong2019arxiv}.  Thus, a simple and transparent illustration of the implications seems to be helpful for individuals to have a complete understanding of differential privacy. 

Considering the abstract nature of the data perturbation process in DP models, especially the \textit{transformation}~\cite{lowe2004interrogation} of the data before and after the noise addition, we also proposed the stepwise illustrations (e.g., animation) to facilitate participants' mental representation process and the comprehension of the illustrations. 

\subsubsection{Privacy-Utility Tradeoff}
Empirical studies have mostly focused on communicating the privacy benefit of differential privacy~\cite{bullek2017towards,xiong2019arxiv}.  
In real world scenarios, users made the data-sharing decisions by evaluating more than one attribute that may influence the final decision~\cite{krause2008utility,luce1964simultaneous}.  Besides privacy benefit, differential privacy introduces utility cost.  Algorithms that follow the concept of DP have a privacy parameter $\epsilon$ that determines the tradeoff between privacy and utility for a request~\cite{zhu2017differentially}.  Given DP, there is a natural tradeoff between information loss and privacy.
Thus, we propose to illustrate privacy-utility tradeoff of the three DP models. 
\section{Overview of Experiment Design}
\label{sec:exp_overview}

We conducted two online surveys ($N=300$ and $N=295$) examining the effects of illustrations in participants’ comprehension of the DP models, their perceived utility and privacy protection, and data-sharing decisions. 
Experiment 1 addresses \textbf{RQ1} and \textbf{RQ2}, and Experiment 2 further addresses \textbf{RQ1} and \textbf{RQ3}. 
A pilot survey and an interview study were conducted before each experiment, evaluating the initial illustration design and survey questions. 
Findings in these studies led to improved illustrations and survey questions that were examined in the experiments. 
We measured participants' data-sharing decisions and noise-level preference in two types of scenarios (public good, commercial interests). 
To contextualize the corresponding decision making, we asked participants to imagine that they were one of the users in the described scenario. 

The illustrations and survey instruments of the experiments can be found in Appendix~\ref{sec:illstruation_i} to \ref{sec:survey_protocol_II}.
The exact description of the scenarios are shown in Appendix~\ref{sec:survey_protocol_II}. 

\subsection{Participant Recruitment}
Both the pilot surveys and the formal experiments were conducted on Amazon Mechanical Turk (MTurk), and the human intelligent tasks (HITs) were posted with restrictions to US workers with at least 95\% approval rate and 100 or more approved HITs.  We made these restrictions in the studies to accurately represent sample restrictions of most recent MTurk research~\cite{hauser2016attentive}.  Participants of the interview study were recruited through emailing acquaintances who had limited knowledge or prior experience with any DP technique. 
All experiments complied with the American Psychological Association Code of Ethics and were approved by the Institutional Review Board (IRB) at the authors' institutes.  Informed consent was obtained for each participant.  Data of the experiments were anonymized before analysis. 

\subsection{Differential Privacy Illustration Design}
\label{sec:design_comm}
To come up with the illustrations of Central DP, Local DP, and Shuffler DP, we started from the data flow descriptions evaluated by~\cite{xiong2019arxiv}, which showed the best comprehension results from end users.  To make the three DP trust models comparable, all the illustrations followed the same style and logic: We first presented a {\it text description}, which was followed by the corresponding \textit{data flow diagram}.  
We expected that the text descriptions would help participants' conceptualize DP when viewing the data flow diagrams. 
A {\it utility heatmap} showing the utility cost at the aggregated level was presented at the end. 

After designing the illustrations, we conducted multiple rounds of internal discussion and review of the illustrations. In the discussions, we involved DP experts to ensure that our illustration
of each model was technically accurate, and laypeople to help ensure that they were easily understandable. Next we describe the illustration of the Shuffler DP model in Experiment 1 (Fig~\ref{fig:diagram_ShufflerDP_all}) as an example. 

\paragraph{Text description} 
Besides describing the DP data flow, we made the implication of the DP model explicit in each text description (e.g., collusion between the shuffler and the server of the shuffler DP results in little benefit of shuffling). 
Based on the pilot survey and the interview study findings, we added a legend listing the set of icons used in the diagram and described the meaning for each of them.  Key icons were also embedded in the text description to help participants associate the text and the diagram (see Appendix~\ref{sec:app_a}, Fig~\ref{fig:central_exp}).  In addition, we improved the wording and emphasized the data perturbation processes and implications for privacy protection.

\paragraph{Data flow diagram}
The data flow diagram starts from the data collection of individual users (see Fig~\ref{fig:diagram_ShufflerDP}).  A snapshot of the map includes a red pin, indicating the actual location of a user.  A gear icon represents the DP technique. After the processing of the DP technique, a user's actual location is blurred with some noise (e.g., it becomes an orange pin at somewhere else) such that the user's presence at the location becomes uncertain.  We vary the noise perturbation across the users. For example, while a single noise obscured user 1's actual location, user $n$'s actual location was replaced by another one.  Then the shuffler is introduced. A security lock is used to indicate an extra layer of security added to the perturbed data in the shuffler database.  Data shuffling (e.g., data of user 1 assigned to user 6) is presented afterward. A green shuffle icon is also presented to indicate the break of the linkage between the users and their data.  After data shuffling, an encryption key is used in the App database to indicate the unlock of the security protection for data publishing to data analysts or collaborators. 

We also improved the diagram based on the results of the pilot survey and the interview.  To increase the contrast in color coding, we used the yellow color referring to the perturbed data with DP protection, and the green color showing the shuffled data after another layer of security protection. For color-deficient participants, we also added a dashed line to code the perturbed data in the data flow (see Fig~\ref{fig:diagram_ShufflerDP}). 
\vspace{-0.2cm}
\begin{figure}[H]
    \centering
    \begin{subfigure}{0.45\textwidth}
         \includegraphics[width=\textwidth]{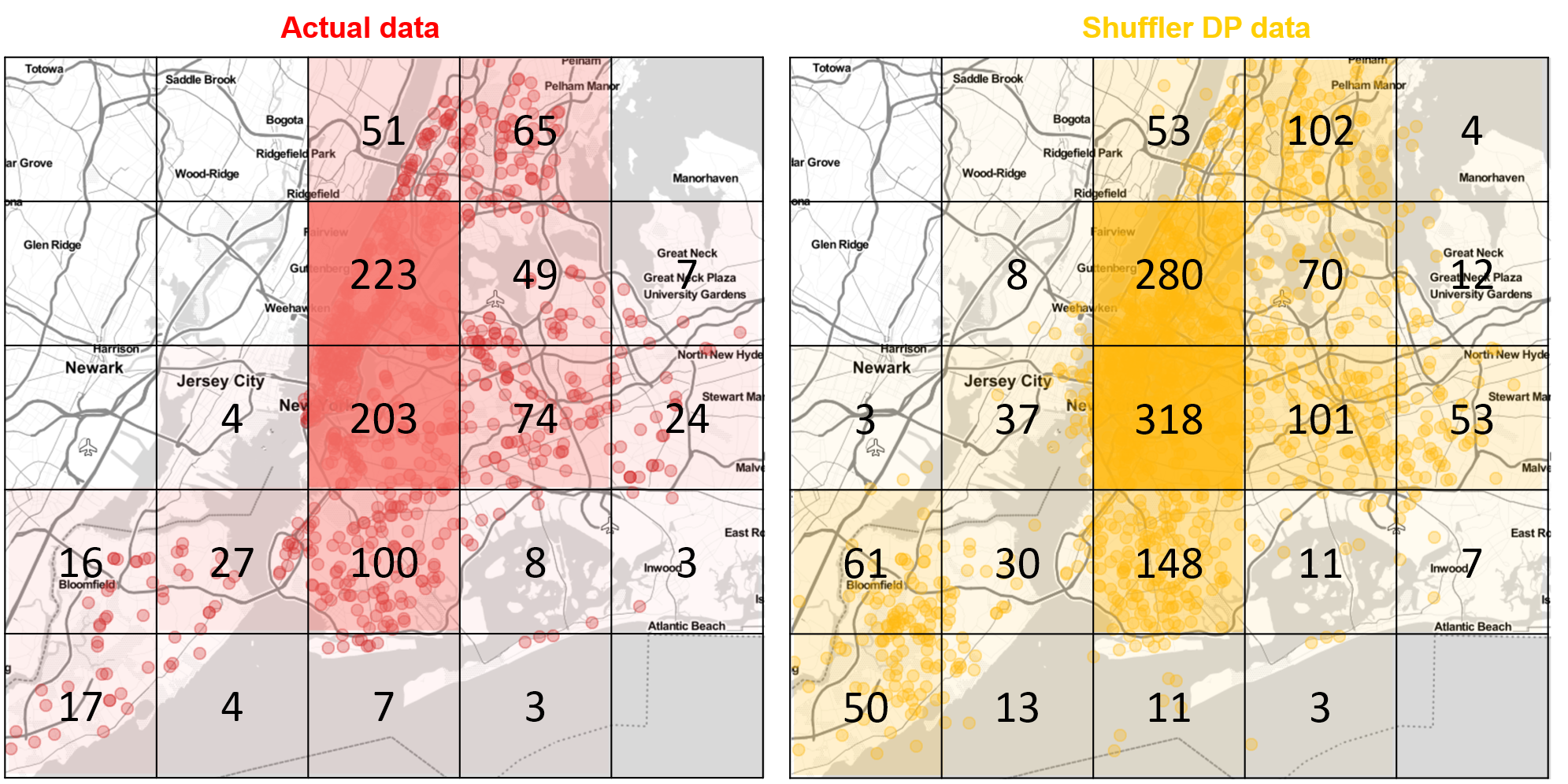}
     \end{subfigure}
        \caption{Illustrations of utility heatmap for the Shuffler DP model evaluated in Experiment 1.}
        \label{fig:shuffler_exp_main}
\vspace{-0.3cm}        
\end{figure}

\paragraph{Utility heatmap}  We also proposed illustrations showing how the DP model impacts the utility of the collected data at an aggregated level by comparing it to actual data before data perturbation (see Fig~\ref{fig:data_ShufflerDP}). Consistent with the data flow diagram, the red dots indicate users' actual locations while the orange dots represent the perturbed location information. 
To enhance the illustration of the utility implication, we added a layered heatmap to the original data visualization and labeled the number of data points in each cell. We used positive noise rather than unbiased noise in DP on purpose to make utility cost easily understandable to laypeople. Fig~\ref{fig:shuffler_exp_main} shows the improved data visualizations for the Shuffler DP, which is also explained with the textual description (see Appendix~\ref{sec:illstruation_i}).

\paragraph{Stepwise illustrations}
In the pilot survey and the interview study, participants tended to skim through the illustrations and omit details. Thus, we divided the illustrations into multiple pages to increase users’ attention to the detailed visualization~\cite{hong2004does}. 
In Experiment 1, we presented the stepwise illustration using animation. 
The materials used in the animation condition were the same as those in the illustration condition except that the graphics and texts were combined and animated as videos to illustrate the data flow step by step. 
The animated videos were further narrated by a native English speaker. The video for each model lasts for 88 seconds (Central DP), 85 seconds (Local DP), and 134 seconds (shuffler DP), respectively. 
To foreshadow, we did not obtain any significant differences between the illustration and the animation conditions. Thus, we implemented the stepwise illustrations without animation in Experiment 2. 

The illustrations for the Central DP and the Local DP were improved in a similar way. Specifically, we emphasized the different levels of noise in the data perturbation processes, the meaning of the model-specific process, as well as the privacy protection implications and utility implications for each model. See Appendix~\ref{sec:illstruation_ii} for a detailed descriptions.

\section{Experiment 1}
\label{sec:exp}

The primary goal of our study was to design and evaluate effective communication of the three DP models (Central DP, Local DP, and Shuffler DP) to end users. After designing the initial illustrations (i.e., text descriptions and graphics) to convey the features of the models, we conducted an online pilot survey ($N=30$) and then more detailed interviews of lay users ($N=6$). 

In summary, the pilot survey and the interview study revealed that the proposed illustrations were not effective in communicating DP models, and some reasons why that was so. First, the distribution of survey time (e.g., the 15-second median viewing time of the illustration of each model) and ``Did not read it carefully'' theme in the interview indicate that participants tended to skim through the pages and omit details. This problem might be alleviated by dividing the current illustrations into multiple pages and presenting the whole with an animation to increase users’ attention to the detailed visualization~\cite{hong2004does}. We expected the reduced information on each page (e.g., dividing the data perturbation process into multiple pages) will help participants comprehend the key aspects. 
The ``Local DP vs. Shuffle DP'' theme in the interview revealed that participants had difficulty understanding the implications of different data perturbation processes. Thus, besides clearly presenting technical details, it is critical to emphasize the data perturbation implications on security and privacy. We also considered that a direct comparison across models may assist users to understand the difference in practice, and impact their data-sharing decisions.

We improved the illustration of each model based on the findings from the pilot survey and the interview study. 
We also generated a stepwise illustration using animation to make the key visualizations about data perturbation more explicit. We conducted a formal online survey study to evaluate how the improved illustrations impact users’ comprehension, perception, and data-sharing decisions across the three DP models. 

\label{sec:survey_pro}

\begin{figure}[ht!]
  \centering
  \includegraphics[width=0.46\textwidth]{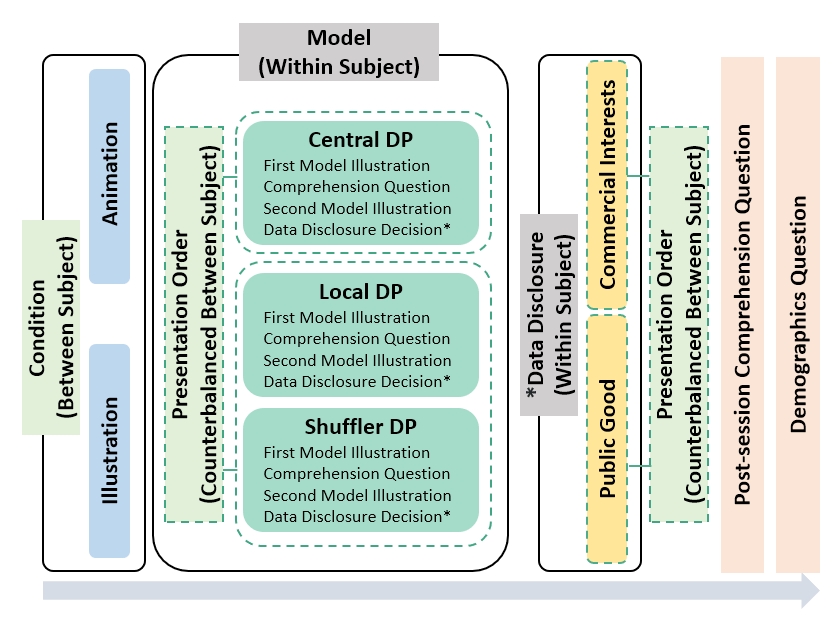}
  \caption{We used a mixed design for Experiment 1, with condition (illustration, animation) between subjects and model (Central DP, Local, DP, Shuffler DP) (\textbf{RQ 1}) and data usage scenario (public good, commercial interests) (\textbf{RQ 2}) within subjects. 
  The flow chart of Experiment 2 was similar except that model (\textbf{RQ 1}) became a between-subject factor. Moreover, we examined participants' preferred noise level (low, medium, high) (\textbf{RQ 3}) for each DP model across the same two scenario categories as Experiment 1.     
  }
  \label{fig:exp_flow}
  \vspace{-0.3cm}
\end{figure}

\subsection{Participants.} We recruited $400$ participants from MTurk. Four duplicate survey responses were removed. We further filtered participants by the survey duration. The median survey completion time was about $20$ min. Since it took $10$ min to watch all videos, we used 
$15$ min as the lower threshold.  
We plotted the distribution of completion time and cut off responses more than an hour.  As a result, we included $300$ participants in the data analysis, with $160$ in the illustration condition and $140$ in the animation condition. Among those $300$ participants, $147$ of them viewed the Central DP at first and $153$ of them viewed the Local DP at the very beginning.  Participants were mostly White ($75.3\%$), slightly more male ($57.0\%$) than female, and most in the age range of $25-44$ years ($76.3\%$). About $79.6\%$ of the participants had a Bachelor’s degree, Medical degree, or Professional degree. About $57.6\%$ of them had experience with computer or information technology field.

\subsection{Stimuli and Procedure}

The survey flow (see Fig~\ref{fig:exp_flow}) was similar to that of the pilot survey and the interview study. Participants were randomly assigned into the illustration condition or the animation condition.  There was a 60 s minimum viewing time for each illustration in the illustration condition. The animation of each model was automatically played.  When each animation ended, participants were directed to the next page automatically to prevent them from being distracted by the links to the YouTube website. After answering the comprehension question following each model (see Appendix~\ref{sec:survey_protocol}), participants received feedback about their performance and were instructed to read/watch the illustration or animation again for the second time.  
A 7-point Likert Scale was used to evaluate the perceived utility and privacy protection, and data-sharing decisions (see Appendix~\ref{sec:survey_protocol_II}). The survey took a median of $20$ min to complete on average, and the payment was \$3.50 for each participant.


\subsection{Results}

Correct answer rate of the comprehension question for each model collapsed across participants (see Table~\ref{tab:comprehension}) were entered into 3 (model: Central, Local, Shuffler) $\times$ 2 (presentation: once, twice) $\times$ 2 (order: Central DP first, Local DP first) Chi-squared tests. Post-hoc tests with Bonferroni corrections~\cite{bland1995multiple} were performed, testing all pairwise comparisons with corrected \textit{p}-values for possible inflation.  Participants’ average rating for perceived usefulness and security/privacy were analyzed with ANOVA using the same three factors as the chi-squared tests, respectively. Participants' data-sharing decisions were entered into 3 (model: Central, Local, Shuffler) $\times$ 2 (usage: research, commerical) $\times$ 2 (order: Central DP first, Local DP first) ANOVA.  Post-hoc tests were also performed for both perception and decision-making measures.


\begin{table}[ht]
\vspace{-0.1cm}
\caption{Correct answer rate for the comprehension question of each DP model (after viewing the illustrations/animations once and twice). Numbers in the parentheses indicate the number of participants in each condition. }
\footnotesize
\centering
\resizebox{0.48\textwidth}{!}{
\begin{tabular}{lcccccccc}
\toprule
 & \multicolumn{2}{c}{Central DP} &  & \multicolumn{2}{c}{Local DP} &  & \multicolumn{2}{c}{Shuffler DP} \\ \cline{2-3} \cline{5-6} \cline{8-9} 
      & Once  & Twice &  & Once  & Twice &  & Once  & Twice \\ \midrule
Illustration (160) & 72.5\% & 79.4\% &  & 45.6\% & 47.5\% &  & 71.9\% & 80.0\% \\
Animation (140) & 64.3\% & 78.6\% &  & 56.4\% & 47.1\% &  & 75.7\% & 79.3\% \\ \bottomrule
\end{tabular}}
\label{tab:comprehension}
\vspace{-0.1cm}
\end{table}

\subsubsection{Comprehension} 
\label{sec:formal_comp}
Table \ref{tab:comprehension} shows the correct answer rate for comprehension questions across the three models after viewing the illustrations or animations once and twice.  Due to the unbalanced number of participants, we did not conduct statistical analysis comparing the results of the formal survey and the pilot survey.  However, the correct rates were numerically higher than those in the pilot survey in general.  Compared with the Central DP and the Shuffler DP, the Local DP had an overall lower correct rate (${\chi}^2_{(2)}=39.83, \, p<.001$). 
The question for Local DP asked about the privacy implication of data sharing with a third party, which was not directly explained in the illustration (see Appendix~\ref{sec:survey_protocol} CQ1-2). The low correct rate suggests that participants may only have grasped information explicitly expressed for the model. 

Comparing the results of the watching animations/illustrations once versus twice, 
only Central DP showed a significant increase (${\chi}^2=7.76, \, p =.005$), and such pattern was more evident for the animation condition than for the illustration condition ($p = .002$). 
We further examined whether the order of model presentation had an effect. 
When comprehension questions were asked for the first time, there was no significant difference between the two presentation orders.  However, for the second time, the correct rate for the Central DP was significantly higher when the Central DP was presented last than when it was presented first (${\chi}^2=9.00, \, p=.003$). Thus, the increased correct rate of the Central DP could be attributed to the order of presentation instead of increased understanding of the model after viewing the illustration again.  When Central DP was presented last, participants could have clearer memory of the corresponding information.  The Local DP was either presented first or second. The Shuffler DP was presented second or last, indicating the initial high correct rate and non significant increase for the second time.

Based on the results of comprehension questions, we further filtered the data for the analysis hereafter. Specifically, only participants with at least two correct answers in either the first or the second time were kept, resulting in 261 responses (138 in the illustration condition, 123 in the animation condition, 129 with Central DP presented first, and 132 with Local DP presented first). 

\begin{figure}[ht]
\includegraphics[width=0.48\textwidth]{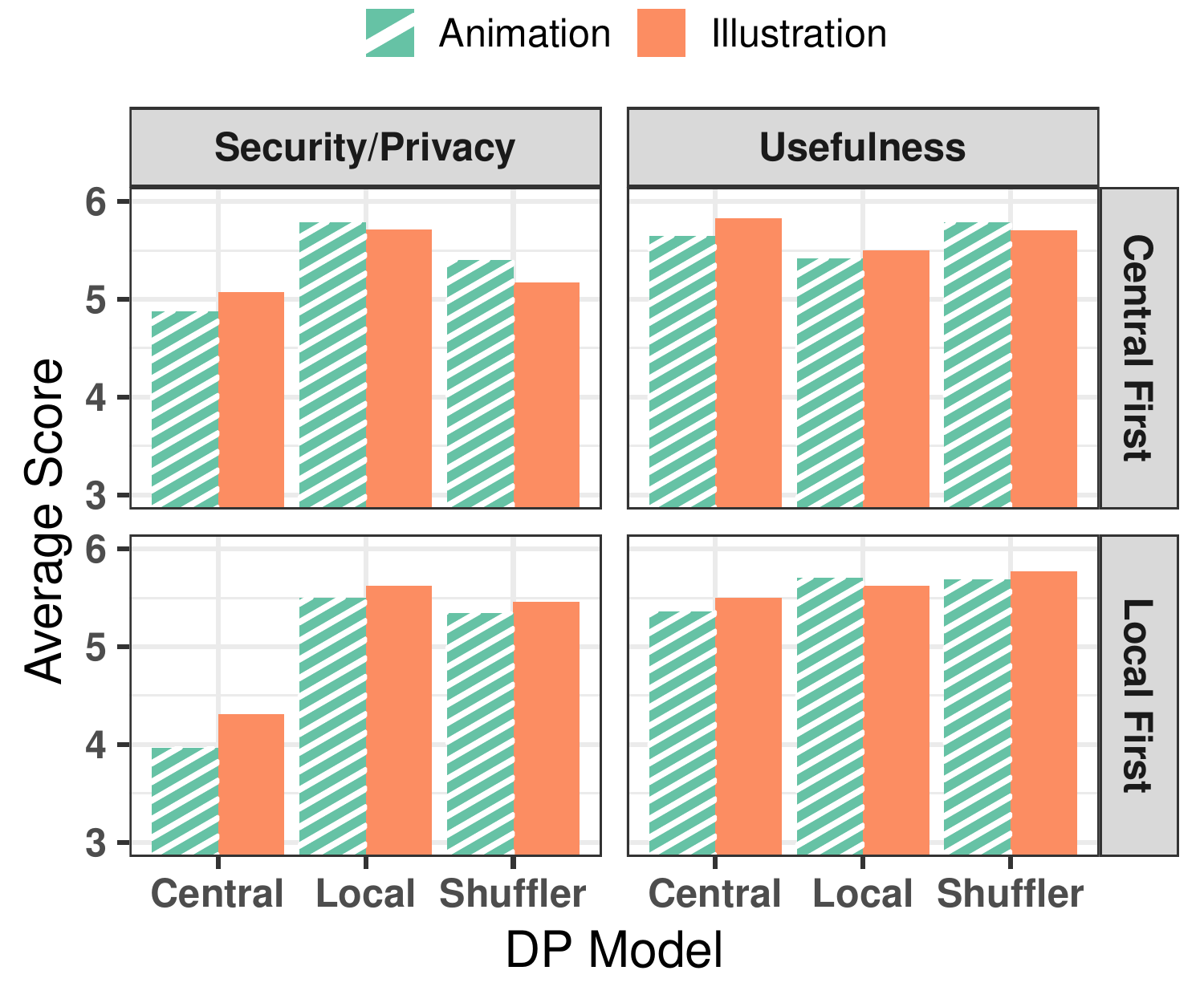}
\vspace{-0.4cm}
\caption{Average score for security/privacy and usefulness ratings as a function of condition (animation, illustration), DP model (Central, Local, Shuffler) and DP model presentation order (Central first, Local first). 
}
\label{fig:perception}
\vspace{-0.1cm}
\end{figure}

\begin{figure}[ht]
\includegraphics[width=0.48\textwidth]{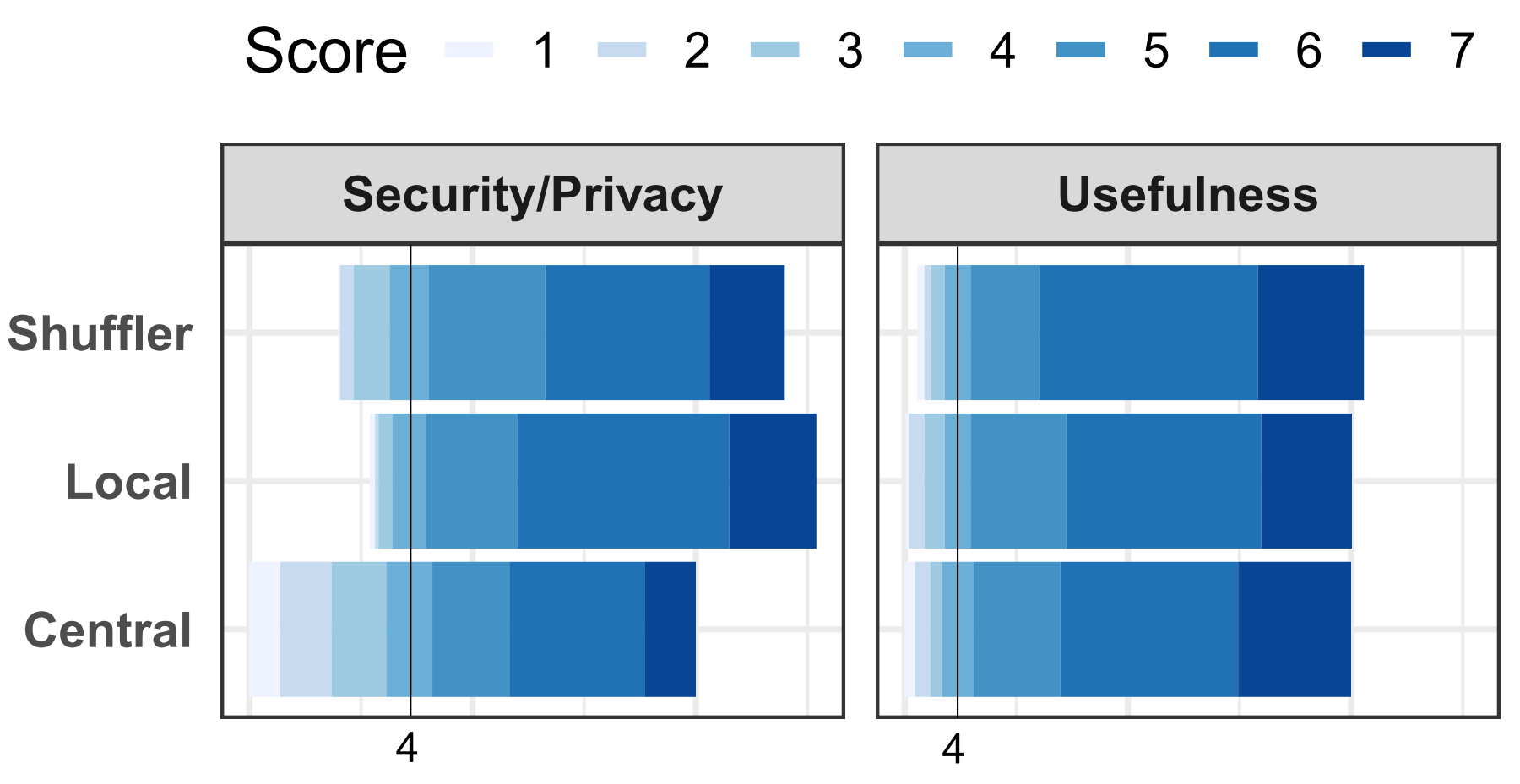}
\vspace{-0.4cm}
\caption{Overall perceived security/privacy and usefulness ratings for the three DP models.}
\label{fig:bar_perception}
\vspace{-0.3cm}
\end{figure}

\subsubsection{Privacy and Utility Perception}
\label{sec:formal_perception}
Fig~\ref{fig:perception} shows participants' average rating of perceived security/privacy and usefulness for the three models. Fig \ref{fig:bar_perception} shows the proportion of participants in each rating score, aligned by 4 (neither disagree nor agree). For the perceived security/privacy, the main effects of model ($F_{(2,771)}=40.75,\, p<.001,\, {\eta}^2_p=.10$), presentation order ($F_{(1,771)}=8.49,\,p=.004,\, {\eta}^2_p=.01$), and their interaction ($F_{(2,771)}=7.65,\, p<.001,\, {\eta}^2_p=.02$), were all significant. Post-hoc comparisons revealed that participants gave higher rating for the Shuffler DP ($5.35$) than that of the Central DP ($4.56,\,p_{adj} < .001$), but the rating of Shuffler DP was lower than that of the Local DP ($5.66,\,p_{adj} =0.04$). Such results demonstrate a correct understanding of privacy implications across the three models, which was not observed in the pilot studies.  Post-hoc comparisons also indicated that the ordering effect was only evident for Central DP. Specifically, when Local DP was presented firstly, the average rating for the Central DP was lower ($4.14$) than when the Central DP was presented initially ($4.98,\,p_{adj} < .001$). 
Thus, the presentation of Local and Shuffler DP could impact people's perceived security/privacy of Central DP, but not vice versa.

In terms of the perceived usefulness, the ANOVA showed no significant effect at all.  Given the heatmap and numbers (see Fig~\ref{fig:shuffler_exp_main}), the utility implications of DP models should not be difficult for participants to understand.  A possible explanation for the obtained results is that the reduced accuracy of the three models were all acceptable for the participants.  Alternatively, participants might be less concerned about the utility aspect compared to the privacy aspect.  


\subsubsection{Data-sharing Decision}
\label{sec:formal_decision}
\begin{figure}[ht]
\includegraphics[width=0.48\textwidth]{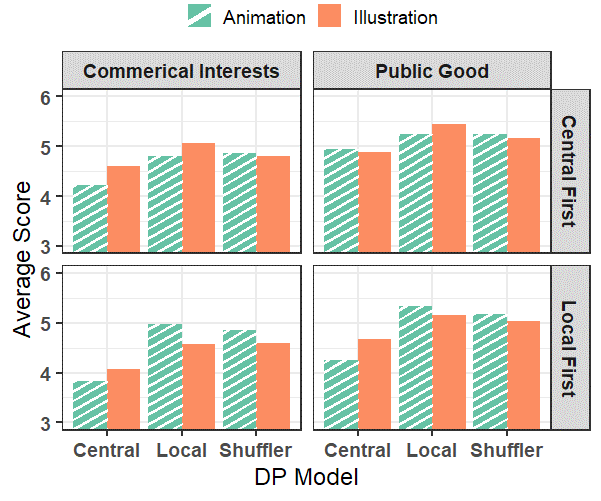}
\vspace{-0.2cm}
\caption{Average score for data disclosure in the public-good and commercial-interests scenarios as a function of condition (animation, illustration), DP model (Central, Local, Shuffler), and DP model presentation order (Central first, Local first). 
}
\label{fig:disclose}
\vspace{-0.1cm}
\end{figure}

\begin{figure}[ht]
\includegraphics[width=0.48\textwidth]{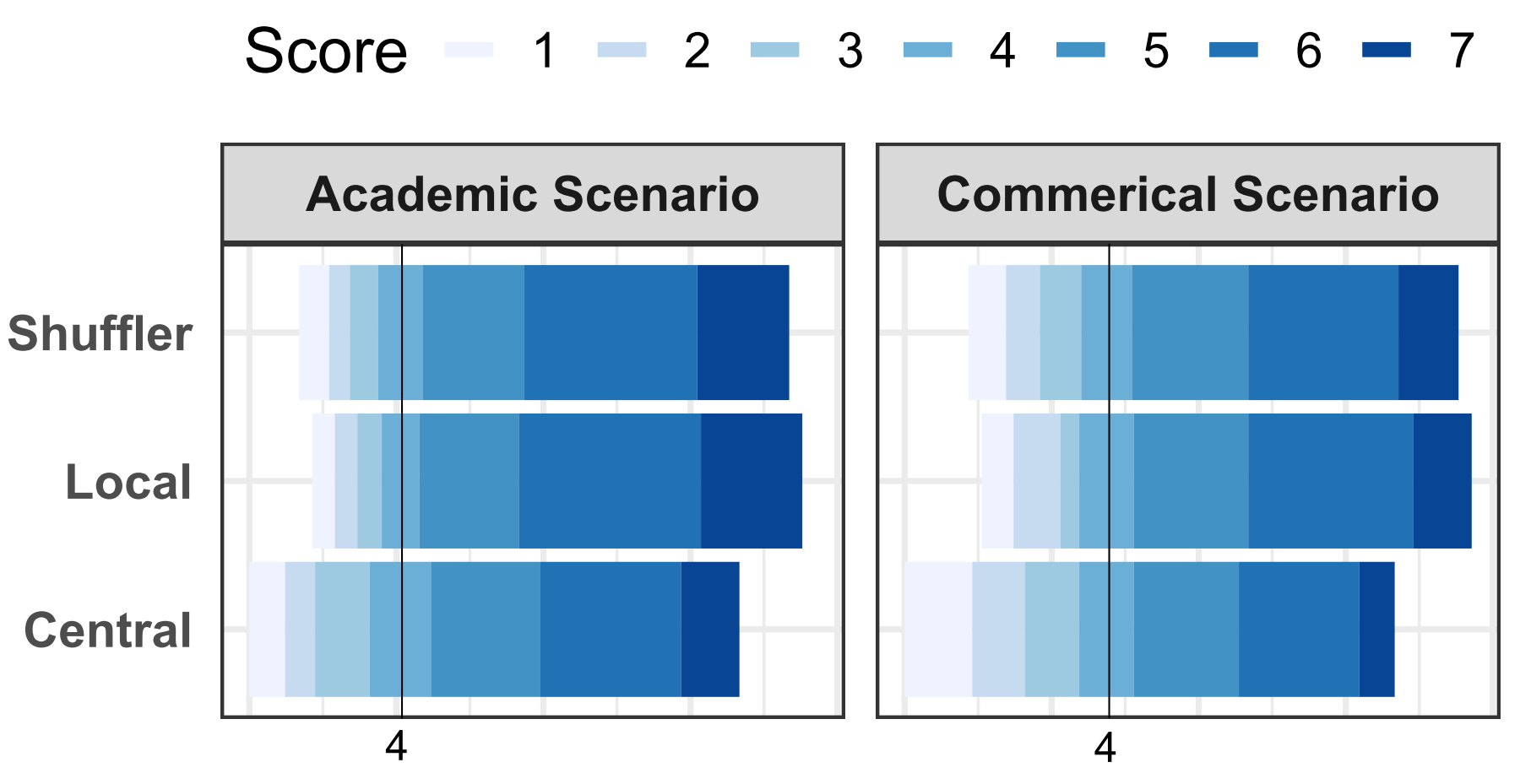}
\caption{Overall data disclosure in the academic (i.e., public-good) and commercial (i.e., commercial interests) scenarios for the three DP models (Central, Local, Shuffler).}
\label{fig:bar_discolse}
\vspace{-0.4cm}
\end{figure}

Fig~\ref{fig:disclose} shows participants' willingness to share location data across the three models. Fig \ref{fig:bar_discolse} shows the proportion of participants in each rating score, aligned by 4. ANOVA showed the main effects of usage scenario ($F_{(1,1542)}=26.51,\,p<.001,\, {\eta}^2_p=.02$), model ($F_{(1,1542)}=19.30,\,p<.001,\, {\eta}^2_p=.02$), and presentation order ($F_{(1,1542)}=7.39,\,p=.007,\, {\eta}^2_p=.01$). Participants were more likely to share their data for research in disease control ($5.05$) than for commercial usage ($4.60,\, p_{adj}<.001$). 
They were more likely to share with the Local DP ($5.06$) or the Shuffler DP ($4.96$) than with the Central DP ($4.44, p_{adjs}<.001$),  
indicating the preference for stronger privacy protection.  Participants were also more likely to share data when the Central DP was presented first, which again implies the impact of the Local DP and the Shuffler DP on participants' evaluation of the Central DP. 

\subsection{Discussion}
In summary, we found that participants preferred stronger privacy protection 
when they were asked about their preferences for sharing location data. This was independent from how the DP model was communicated or the type of data usage. Users in both conditions preferred the Local DP or the Shuffler DP to the Central DP for the public-good or the commercial-interests scenarios. Lastly, the model presentation order influenced participants' privacy protection perception and the data-sharing decisions: Users showed less perceived privacy protection and preference to share data with the Central DP when it was presented after the Local DP and the Shuffler DP.  The benefit of data sharing or the utility aspect seems to be weighted less than the privacy aspect by the participants. 
\section{Experiment 2}
\label{sec:exp2}
Experiment 2 was conducted to further examine participants' privacy protection and utility cost of the three models using a between-subject design (\textbf{RQ1}). 
Moreover, we introduced the concept of noise level and measured participants' preference of the noise level for different data usage scenarios (\textbf{RQ3}). Another interview study ($N=9$) and additional pilot survey ($N=29$) were conducted before the formal study. The interviews identified problems about the definition clarity, instruction clarity and consistency issues in the survey instrument and model illustrations. After address these problems, we conducted the pilot study and found that that participants could complete the survey with a reasonable completion time and comprehension of DP models. 

\subsection{Participants}
Participants were recruited through MTurk. 
A total of $368$ valid survey responses were collected. 
Each participant was paid for \$4.
We removed one duplicate response. 
The median completion time was $13$ min, which did not differ across the three models. Considering that the overall completion time was not long, we applied half of the median completion time of each model as a lower threshold (Central DP: 6 min, Local DP: 6.5 min, and Shuffler DP: 7.25 min). As a result, there were $99$ participants in the Central DP condition, $90$ in the Local DP condition, and $106$ in the Shuffler DP condition. 

The demographic information shows a similar pattern to Experiment 1. Participants were mostly White ($72.2\%$), slightly more male ($56.3\%$) than female, and most in the age range of 25-44 years ($69.2\%$). About $82.0\%$ of the participants had a Bachelor’s degree, Medical degree, or Professional degree. About $61.4\%$ of them had experience with computer or information technology field. 


\subsection{Stimuli}

\subsubsection{Differential Privacy Illustration}
The illustrations of the three DP models were the same as Experiment 1 except that the wording and formatting were modified to improve the clarity and readability based on the results of the second pilot survey and the second interview study. The modified illustrations can be found in Appendix~\ref{sec:illstruation_ii}. Moreover, we added two extra stepwise illustrations, \textit{Privacy Risks} and \textit{Noise Level}. 

\textit{Privacy Risks Illustration.}
To enhance participants' comprehension of the privacy protection from DP, we added the illustrations of the privacy risks before the model illustration. 
Specifically, we set up a scenario where users' locations collected through installed apps were anonymized and then published as an aggregated map. 
The map visualization was generated based on an open dataset provided by NYC Open Data.\protect\footnotemark   
  We then explained how an anonymized individual user in the grid map could be easily identified than others (see details in Appendix~\ref{sec:survey_protocol_II}).

\footnotetext{ https://data.cityofnewyork.us/Transportation/Parking-Meters-GPS-Coordinates-and-Status/5jsj-cq4s}


\textit{Noise Level Illustration.}
After the model illustration, we further explained the concept of noise in DP and how an increased  level of noise can enhance privacy protection but reduce data accuracy. Participants were required to experience three noise levels (i.e., low, moderate, and high) through clicking corresponding level buttons. Upon clicking, the visualization of the aggregated map corresponding to each level of noise was shown to the participants (see Fig \ref{fig:noise_level} in Appendix.\ref{sec:survey_protocol_II} 

\subsubsection{Comprehension Questions}
Comprehension questions are critical for evaluating participants' understanding of DP mechanisms. Based on questions used in Experiment 1, we constructed new questions and improved the questions based on the second pilot survey and the second interview study.
In each model, we presented CQ2-1 after the privacy risks illustration and tested whether participants understand that anonymization cannot effectively protect user privacy. CQ2-2 examined participants' comprehension of when noise is added to the data flow. Thus, the correct answer was model-dependent. Participants answered CQ2-2 after the data-flow illustration. CQ2-3 was exclusive to the Shuffler DP and examined whether participants understood the function of the shuffling process. CQ2-4 and CQ2-5 were presented after the noise level illustration, testing whether participants comprehended how different noise levels impact data privacy and accuracy (e.g., increasing the level of noise for better privacy and decreasing the noise level for better data accuracy).
The full descriptions of comprehension questions can be found in Appendix~\ref{sec:survey_protocol_II}.

\subsubsection{Data-sharing Scenarios}
To understand participants' preferences for the level of noise in DP, we constructed four data-sharing scenarios (see Appendix~\ref{sec:survey_protocol_II}) based on the two same categories (public good, commercial interests) as Experiment 1. In the two public-good scenarios, 
we asked participants' to imagine that their and other car owners' parking locations will be collected and used for decision-making related to building electrical vehicle charging stations or identifying sufficient disabled parking lots. In contrast, participants were asked to imagine that the same data collection and use for commercial advertisements or parking garage investment in the commercial-interests category. For each scenario, before making the noise level selection, participants were required to view the data visualization of the three noise levels (low, moderate, and high) again.

\subsection{Procedure}
The survey flow 
was similar to that of Experiment 1 except as noted. First, only one of three models was randomly assigned to each participant. Comprehension questions were presented after the corresponding stepwise illustrations. After completing all comprehension questions, participants rated whether the privacy protection and data accuracy provided by DP met their expectations on a 7-point Likert Scale (``1'' means ``Strongly disagree"; and ``7'' means ``Strongly agree"). The four data-sharing scenarios were then randomly presented to the participants. 
After viewing each scenario, participants were asked to specify their preferred noise level and briefly explain their rationale with an open-ended question.

\subsection{Result}


\subsubsection{Comprehension}
\label{sec:formal_comp_2}
Table \ref{tab:comprehension_ii} shows the correct answer rate of each comprehension question across the three models. 
Across the questions, the correct answer rates were similar among the three models (${\chi}^2_s \leq 1.86$, $p_{s} \geq .440$) except that 
the correct rate was noticeably lower for CQ2-2 in the Central DP model (${\chi}^2_{(2)} = 72.2$, $p < .001$). 
While participants revealed reasonably correct answer rates for CQ2-1 and CQ2-5, their correct answer rates for CQ2-3 and CQ2-4 were worse than chance. 
Thus, we compared the questions and options between those two groups. We found that
the correct answer for CQ2-1 and CQ2-5 was the first option. However, it was not the case for CQ2-2 Central DP, CQ2-3, or CQ2-4. 
While the presented illustrations could not be very effective in helping participants understand DP, we conjecture that the obtained results might be impacted by inattentive or bot respondents on Amazon MTurk~\cite{hauser2016attentive,kennedy2020shape}.  Consequently, we identified $73$ participants who invariably selected the first option for all the comprehension questions.

\begin{table}[]
\caption{Correct answer rate for the comprehension question of each DP model. Numbers in the parentheses indicate the number of participants in each condition.
}
\centering
\resizebox{0.48\textwidth}{!}{
\begin{tabular}{clccc}
\toprule
 &  & \multicolumn{1}{l}{Central DP (99)} & \multicolumn{1}{l}{Local DP (90)} & \multicolumn{1}{l}{Shuffler DP (106)}  \\ \hline
Privacy Risk                 & CQ2-1 & 90.9\% & 92.6\% & 87.9\%            \\
\multirow{2}{*}{DP}          & CQ2-2 & 28.7\% & 79.5\% & 68.6\%  \\
                             & CQ2-3 &   NA      &   NA      & 44.4\%            \\
\multirow{2}{*}{Noise Level} & CQ2-4 & 49.5\% & 40.9\% & 47.6\%            \\
                             & CQ2-5 & 66.4\% & 71.3\% & 69.4\%           \\ \hline
\end{tabular}}
\label{tab:comprehension_ii}
\end{table}

Thus, we further filtered out those $73$ participants. 
Moreover, same as Experiment 1, we only kept participants who correctly answered at least two of the four questions (excluding CQ2-3) that are applicable to all models. For the remaining $217$ participants, there were $62$ in the Central DP condition, $70$ in the Local DP condition, and $85$ in the Shuffler DP condition.


\subsubsection{Privacy and Utility Perception}
\label{sec:formal_percp_2}
Regarding whether the DP model meets participants' expectations for privacy, 
participants' average ratings were $5.92$ (Central DP), $5.77$ (Local DP), and $5.76$ (Shuffler DP), all of which showed no significant differences ($F < 1.0$).
Likewise, they gave similar perceived data accuracy across the models ($F < 1.0$): $5.55$ (Central DP), $5.84$ (Local DP), and $5.85$ (Shuffler DP). 

\subsubsection{Noise Preference}
\label{sec:formal_noise_2}
Across the two scenario categories, the average proportions of participants opting for low-, moderate-, and high-levels of noise were 32.4\%, 32.8\%, and 32.3\% (${\chi}^2_{(2)}<1.0$) 
The main effect of scenario category was significant (${\chi}^2 =12.89,\, p= .001$). 
Specifically, participants were more likely to select high-level noise in the commercial-interests scenarios (40.19\%) than the public-good scenarios (29.08\%). 
Nonetheless, the effect of scenario category showed no significant differences across the three models (${\chi}^2_{(2)}=4.11,\, p=.392$). 

\subsubsection{Responses to Open-ended Questions}
\label{sec:formal_open}
We asked participants to briefly explain why they selected the specific noise level. 
After disregarding meaningless responses, such as ``good ($97$)'' and ``noise is unwanted sound ($53$).'' 
Collapsed across the three models, we conducted a thematic analysis~\cite{braun2006using} using the remaining $212$ meaningful responses from $53$ participants. 
One co-author and a graduate student working in another co-author's lab performed the thematic analysis independently at first. Then, they discussed the results and finalized the thematic analysis together. We identified four major themes of participants’ responses as follows. 

 {\it Strong Protection for the Commercial-interests Scenarios.} Among the $212$ answers, 30\% of the participants described that they ``don’t like (P$1$)'' or ``don’t care about (P$13$)” advertisements. They also believed that those companies ``... already have too much data on everyone (P$27$).'' Thus, they ``would much rather have the highest amount of data protection (P$12$)'' in those scenarios.  
 
{\it Correct Comprehension of DP Models.}
About 22\% of the participants revealed somewhat comprehension of the DP models in their responses. For example, participants described that they considered both ``keeping the accuracy of data (P$2$)'' and ``protecting user privacy (P$21$)'' when making the decisions. The tradeoff between privacy and accuracy may have increased participants' preferences for the moderate level. For example, among the participants ($12$) who discussed the tradeoff in one of the public-good scenarios, 50\% opted for the moderate level, 34\% for the low-level, 8\% for the high level, and 8\% chose the ``Unsure'' option.

{\it Individual Differences for the Public-good Scenarios.}
About 19\% of the participants selected either the low-level or the moderate-level noise for the public-good scenarios, e.g., disabled parking.  Participants explained that data accuracy ``is highly important (P$11$)'' and they would like to ``make sure that there is always an adequate amount of disabled parking (P$3$)''. In contrast, about 10\% of the participants opted for the high-level noise and described that ``disability is very private and needs the maximum amount of privacy (P$37$).''

{\it Individual Differences Regardless Scenarios.} Moreover, $4$ (8\%) participants chose the high-level noise across the scenarios. They indicated that they ``less concerned about the accuracy of the data (P$30$).'' Instead, they described that ``vehicle location is very sensitive (P$41$)''  and ``you never know who will access (the) data (P$23$).'' Consequently, they argued that ``high level should be deployed at all times (P$41$)''. Similarly, another $4$ (8\%) of the participants selected the low-level and the moderate-level, respectively. They believed that the selected level ``would probably be good enough (P$50$)'' to prevent identification, and ``somewhat accurate data is still available (P$12$).''

\subsection{Discussion}
Experiment 2 employed procedures similar to Experiment 1 except for the between-subject design. 
Using a new set of comprehension questions, we obtained similar results as Experiment 1. Moreover, we found the results might have been impacted by inattentive respondents, which we discuss in the General Discussion. 
The effect of model was significant for the perceived privacy in Experiment 1. Thus, the non-significant results in Experiment 2 suggested that the obtained results in Experiment 1 are mainly due to the within-subject design (i.e., the relative comparisons across the three models). When only one DP model was presented, participants seemed to give high ratings of perceived privacy and data accuracy in general.
Moreover, participants preferred the high-level noise protection for the commercial-interests scenarios than for the public-good scenarios, in agreement with the same effect at the model level in Experiment 1.

\section{General Discussion}
In this work, we conducted two online experiments (each was proceeded by a pilot survey and an interview study) examining participants' comprehension and perceptions on privacy and utility of three differential privacy (DP) models, and their data-sharing decisions.  As relevant factors, we investigated three DP trust models (Central, Local, and Shuffler), and two data-usage scenarios (public good, commercial interests). 
There are several key findings:
\begin{itemize}[leftmargin=*]
    \item Participants prefer stronger privacy protection at both the model level and noise level.
    \item Participants accept the Shuffler DP model for data disclosure. 
    \item An adequate comprehension of DP is necessary for accurate perception of the privacy protection of different DP trust models and consequent informed decisions.
    \item Compared to the illustration, using animation does not facilitate users' understanding of the models, their privacy and utility perception, as well as their data-sharing decisions.
\end{itemize}
Despite prior work that indicated people may not understand DP procedure~\cite{oberski2020differential}, our work revealed that people \textit{can} comprehend different DP models using illustrations. 
However, those findings must be interpreted with an elaboration on some details.



\subsection{Less Than Ideal Comprehension Performance} In Experiment 1, we asked participants that if the Local DP model is implemented, whether the third party with which the app shared data can see the real answer.  Since the random noise is added at an individual level for the Local DP, the correct response should be ``No.''  However, participants showed a poor comprehension of the privacy implication of the Local DP. In Experiment 2, we obtained similar results for the another four comprehension questions.  Through comparisons, we found a common pattern across those questions: the correct answer was not the first option.  Thus, we identified $16$ participants who always selected the first option for comprehension questions in Experiment 1. We exclude those participants and re-ran the statistical analysis again. The overall results were the same as shown in Experiment 1.

While the obtained results might have been impacted by inattentive respondents on Amazon MTurk~\cite{hauser2016attentive}, the abovementioned results of Experiment 1 indicate that the data filtering criteria (e.g., completion time and correct answer rate for comprehension questions) seemed to be appropriate to exclude invalid responses. 
Nevertheless, there were more inattentive responses in Experiment 2 ($73$) than in Experiment 1 ($16$), revealing the emergent issue of data quality on Amazon MTurk~\cite{kennedy2020shape}.
A second and more likely explanation is to compare the comprehension results of the Central DP.  
In Experiment 1, we asked participants whether an attacker could see the actual location information submitted by them if an attacker got access to the database of the app. We explicitly explained such implications in the text description, and the correct answer rate was about 73\%. In contrast, we asked participants whether the initial data received by an app company contained any noise in Experiment 2. The correct answer rate was about 30\%.  Altogether, the results from both experiments indicate that participants seemed to grasp the information conveyed in the illustrations but failed to infer the implications.

\subsection{Comparisons across the Three DP Models}
\paragraph{Local DP vs. Shuffler DP} Participants perceived stronger security/privacy protection of the Local DP model than that of the Shuffler DP model in Experiment 1 but not Experiment 2. 
One possible reason is due to the within-subject design in Experiment 1, which afforded the relative comparisons across the three models. However, each participant only viewed one DP model in Experiment 2, and gave high ratings of perceived privacy and data accuracy in general. 
Thus, to help users make informed disclosure decision, it seems to be critical to make different DP trust models available for comparison. 
Moreover, participants preferred stronger protection for the commercial-interests scenarios than for the public-good scenarios, in agreement with the same effect at the model level in Experiment 1.

\paragraph{Order effect of Central DP} We obtained the order effect of the Central DP model in both perceived privacy/security and data-sharing decision measures.  Specifically, participants perceived less privacy protection and showed less willingness to share data when the Central DP model was presented after the Local DP and the Shuffler DP models. Since the latter two models provide stronger privacy protection than the Central version, those results indicate participants' preference for strong privacy protection.  Moreover, the order effect of Central DP can also be interpreted as an effect of reference frames~\cite{tversky2000choices}, suggesting that it is essential to evaluate differential privacy from a cognitive perspective.

\paragraph{Similar Perceived Utility across the Three Trust Models} We did not obtained any perceived utility differences across models or conditions in both experiments. Thus, such results were evident at the model level (e.g., qualitative) and the noise level (e.g., quantitative). 
Compared to the privacy aspect, the benefit of data sharing seem to be less critical to the participants. 

\subsection{(In)Effectiveness of the Illustrations} 
Our findings indicate that the explanative illustrations are effective in communicating DP. 
Meanwhile, we also obtained results pointing toward directions for further improvements. 
Our results showed that animation did not add any benefits compared to the static illustrations.  The online survey was conducted remotely with participants' own devices. Various factors, such as volume and background, might have impacted the effect of the animation. Previous studies also showed that animation with oral commentary did not get better comprehension scores than those who studied equivalent static graphics with written text~\cite{betrancourt2005animation}.  That we did not obtain the benefits of the animation could due to the accompanying text providing all the critical information~\cite{catrambone2002using}.  Due to the prevention of  participants from clicking external video link, we played the animations in an automatic manner.  However, interactivity (i.e., giving control over the space and direction of the animation) has been shown to be a key factor for the effectiveness of animation~\cite{mayer2001learning}. Not only such simple control gives learners time to integrate information before proceeding to the next frame, but it also segments the animation into relevant chunks to facilitate learning.  To further understand the effect of animation, future work could consider allowing participants to control the animation.

\subsection{Limitations and Future Work}

First of all, we recruited MTurk workers in the pilot and formal surveys. Thus, participants are younger, more technical, and more privacy-sensitive than the overall U.S. population~\cite{kang2014privacy}. This is evident in our results, which demonstrate a large percentage of participants have experience with the fields of computer or information technology. We believe these limitations are acceptable, as the public has limited knowledge on differential privacy in general~\cite{oberski2020differential}. Secondly, 
we only asked participants' data-sharing decisions on two data usage scenarios (i.e., public good and commercial interests), which we considered to be reasonably representative.  Thirdly, we did not consider more recent developed DP models~\cite{bohler2020secure,bohler2020secureccs,chowdhury2019outis,roth2019honeycrisp,roth2020orchard}. Future work could consider more diverse usages and latest DP models to validate our findings by performing a replication study. 
Also, to take advantage of the benefit of interactivity, we could consider continuously varied $\epsilon$ values and encourage users to manipulate the parameter to further simulate the effect of random noise, which could improve their comprehension of data perturbation of DP.

{
    \bibliographystyle{abbrv}
	\bibliography{bibs/abbrev0,bibs/Nh,bibs/privacy,bibs/password}

\begin{thebibliography}{10}

\bibitem{uscensus}
J.~M. Abowd.
\newblock Protecting the confidentiality of america’s statistics: Adopting
  modern disclosure avoidance methods at the census bureau.
\newblock
  \url{https://www.census.gov/newsroom/blogs/research-matters/2018/08/protecting_the_confi.html},
  2018.

\bibitem{andres2013geo}
M.~E. Andr{\'e}s, N.~E. Bordenabe, K.~Chatzikokolakis, and C.~Palamidessi.
\newblock Geo-indistinguishability: Differential privacy for location-based
  systems.
\newblock In {\em Proceedings of the 2013 ACM SIGSAC Conference on Computer \&
  Communications Security}, pages 901--914, 2013.

\bibitem{ABCP13}
M.~E. Andr{\'e}s, N.~E. Bordenabe, K.~Chatzikokolakis, and C.~Palamidessi.
\newblock Geo-indistinguishability: Differential privacy for location-based
  systems.
\newblock In {\em CCS}, pages 901--914, 2013.

\bibitem{apple-dp}
Apple.
\newblock Apple\;differential\;privacy\;team, learning with privacy at scale,
  2017.
\newblock Available at
  \url{https://machinelearning.apple.com/docs/learning-with-privacy-at-scale/appledifferentialprivacysystem.pdf}.

\bibitem{balle2019privacy}
B.~Balle, J.~Bell, A.~Gasc{\'o}n, and K.~Nissim.
\newblock The privacy blanket of the shuffle model.
\newblock In {\em Annual International Cryptology Conference}, pages 638--667.
  Springer, 2019.

\bibitem{barkhuus2003location}
L.~Barkhuus and A.~K. Dey.
\newblock Location-based services for mobile telephony: a study of users'
  privacy concerns.
\newblock In {\em 9TH IFIP TC13 International Conference on Human-Computer
  Interaction (INTERACT)}, volume~3, pages 702--712. Citeseer, 2003.

\bibitem{benisch2011capturing}
M.~Benisch, P.~G. Kelley, N.~Sadeh, and L.~F. Cranor.
\newblock Capturing location-privacy preferences: quantifying accuracy and
  user-burden tradeoffs.
\newblock {\em Personal and Ubiquitous Computing}, 15(7):679--694, 2011.

\bibitem{beresford2003location}
A.~R. Beresford and F.~Stajano.
\newblock Location privacy in pervasive computing.
\newblock {\em IEEE Pervasive Computing}, 2(1):46--55, 2003.

\bibitem{betrancourt2005animation}
M.~Betrancourt.
\newblock The animation and interactivity principles in multimedia learning.
\newblock {\em The Cambridge Handbook of Multimedia Learning}, pages 287--296,
  2005.

\bibitem{bittau2017prochlo}
A.~Bittau, {\'U}.~Erlingsson, P.~Maniatis, I.~Mironov, A.~Raghunathan, D.~Lie,
  M.~Rudominer, U.~Kode, J.~Tinnes, and B.~Seefeld.
\newblock Prochlo: Strong privacy for analytics in the crowd.
\newblock In {\em Proceedings of the 26th Symposium on Operating Systems
  Principles}, pages 441--459, 2017.

\bibitem{bland1995multiple}
J.~M. Bland and D.~G. Altman.
\newblock Multiple significance tests: the bonferroni method.
\newblock {\em BMJ}, 310(6973):170, 1995.

\bibitem{bohler2020secure}
J.~B{\"o}hler and F.~Kerschbaum.
\newblock Secure multi-party computation of differentially private median.
\newblock In {\em 29th $\{$USENIX$\}$ Security Symposium ($\{$USENIX$\}$
  Security 20)}, pages 2147--2164, 2020.

\bibitem{bohler2020secureccs}
J.~B{\"o}hler and F.~Kerschbaum.
\newblock Secure multi-party computation of differentially private heavy
  hitters.
\newblock In {\em CCS}, 2021.

\bibitem{braun2006using}
V.~Braun and V.~Clarke.
\newblock Using thematic analysis in psychology.
\newblock {\em Qualitative Research in Psychology}, 3(2):77--101, 2006.

\bibitem{bullek2017towards}
B.~Bullek, S.~Garboski, D.~J. Mir, and E.~M. Peck.
\newblock Towards understanding differential privacy: When do people trust
  randomized response technique?
\newblock In {\em Proceedings of the 2017 CHI Conference on Human Factors in
  Computing Systems}, pages 3833--3837. ACM, 2017.

\bibitem{catrambone2002using}
R.~Catrambone and A.~F. Seay.
\newblock Using animation to help students learn computer algorithms.
\newblock {\em Human Factors}, 44(3):495--511, 2002.

\bibitem{chan2012optimal}
T.-H.~H. Chan, E.~Shi, and D.~Song.
\newblock Optimal lower bound for differentially private multi-party
  aggregation.
\newblock In {\em ESA}, pages 277--288. Springer, 2012.

\bibitem{cheu2018distributed}
A.~Cheu, A.~Smith, J.~Ullman, D.~Zeber, and M.~Zhilyaev.
\newblock Distributed differential privacy via shuffling.
\newblock In {\em Annual International Conference on the Theory and
  Applications of Cryptographic Techniques}, pages 375--403. Springer, 2019.

\bibitem{chow2011trajectory}
C.-Y. Chow and M.~F. Mokbel.
\newblock Trajectory privacy in location-based services and data publication.
\newblock {\em ACM Sigkdd Explorations Newsletter}, 13(1):19--29, 2011.

\bibitem{clark1991dual}
J.~M. Clark and A.~Paivio.
\newblock Dual coding theory and education.
\newblock {\em Educational Psychology Review}, 3(3):149--210, 1991.

\bibitem{cummings2021need}
R.~Cummings, G.~Kaptchuk, and E.~M. Redmiles.
\newblock " i need a better description": An investigation into user
  expectations for differential privacy.
\newblock In {\em Proceedings of the 2021 ACM SIGSAC Conference on Computer and
  Communications Security}, pages 3037--3052, 2021.

\bibitem{nips:DingKY17}
B.~Ding, J.~Kulkarni, and S.~Yekhanin.
\newblock Collecting telemetry data privately.
\newblock In {\em Proceedings of the 31st International Conference on Neural
  Information Processing Systems}, pages 3574--3583, 2017.

\bibitem{duchi2013local}
J.~C. Duchi, M.~I. Jordan, and M.~J. Wainwright.
\newblock Local privacy and statistical minimax rates.
\newblock In {\em 2013 IEEE 54th Annual Symposium on Foundations of Computer
  Science}, pages 429--438. IEEE, 2013.

\bibitem{Dwo06}
C.~Dwork.
\newblock Differential privacy.
\newblock In {\em International Colloquium on Automata, Languages, and
  Programming}, pages 1--12. Springer, 2006.

\bibitem{erlingsson2019amplification}
{\'U}.~Erlingsson, V.~Feldman, I.~Mironov, A.~Raghunathan, K.~Talwar, and
  A.~Thakurta.
\newblock Amplification by shuffling: From local to central differential
  privacy via anonymity.
\newblock In {\em Proceedings of the Thirtieth Annual ACM-SIAM Symposium on
  Discrete Algorithms}, pages 2468--2479. SIAM, 2019.

\bibitem{rappor}
{\'U}.~Erlingsson, V.~Pihur, and A.~Korolova.
\newblock Rappor: Randomized aggregatable privacy-preserving ordinal response.
\newblock In {\em Proceedings of the 2014 ACM SIGSAC Conference on Computer and
  Communications Security}, pages 1054--1067. ACM, 2014.

\bibitem{evfimievski2004privacy}
A.~Evfimievski, R.~Srikant, R.~Agrawal, and J.~Gehrke.
\newblock Privacy preserving mining of association rules.
\newblock {\em Information Systems}, 29(4):343--364, 2004.

\bibitem{fawaz2014location}
K.~Fawaz and K.~G. Shin.
\newblock Location privacy protection for smartphone users.
\newblock In {\em Proceedings of the 2014 ACM SIGSAC Conference on Computer and
  Communications Security}, pages 239--250, 2014.

\bibitem{fisher2012short}
D.~Fisher, L.~Dorner, and D.~Wagner.
\newblock Short paper: location privacy: user behavior in the field.
\newblock In {\em Proceedings of the Second ACM Workshop on Security and
  Privacy in Smartphones and Mobile Devices}, pages 51--56, 2012.

\bibitem{uniqueid}
Google.
\newblock Google chrome privacy notice, 2020.
\newblock Available at \url{https://www.google.com/chrome/privacy/}.

\bibitem{hauser2016attentive}
D.~J. Hauser and N.~Schwarz.
\newblock Attentive turkers: Mturk participants perform better on online
  attention checks than do subject pool participants.
\newblock {\em Behavior Research Methods}, 48(1):400--407, 2016.

\bibitem{hong2004does}
W.~Hong, J.~Y. Thong, and K.~Y. Tam.
\newblock Does animation attract online users’ attention? the effects of
  flash on information search performance and perceptions.
\newblock {\em Information Systems Research}, 15(1):60--86, 2004.

\bibitem{tversky2000choices}
D.~Kahneman and A.~Tversky.
\newblock Choices, values, and frames.
\newblock In {\em Handbook of the Fundamentals of Financial Decision Making:
  Part I}, pages 269--278. World Scientific, 2013.

\bibitem{kang2014privacy}
R.~Kang, S.~Brown, L.~Dabbish, and S.~Kiesler.
\newblock Privacy attitudes of mechanical turk workers and the us public.
\newblock In {\em Tenth Symposium on Usable Privacy and Security ($\{$SOUPS$\}$
  2014)}, pages 37--49, 2014.

\bibitem{kasiviswanathan2011can}
S.~P. Kasiviswanathan, H.~K. Lee, K.~Nissim, S.~Raskhodnikova, and A.~Smith.
\newblock What can we learn privately?
\newblock {\em SIAM Journal on Computing}, 40(3):793--826, 2011.

\bibitem{kennedy2020shape}
R.~Kennedy, S.~Clifford, T.~Burleigh, P.~D. Waggoner, R.~Jewell, and N.~J.
  Winter.
\newblock The shape of and solutions to the mturk quality crisis.
\newblock {\em Political Science Research and Methods}, 8(4):614--629, 2020.

\bibitem{krause2008utility}
A.~Krause and E.~Horvitz.
\newblock A utility-theoretic approach to privacy and personalization.
\newblock In {\em Proceedings of the 23rd National Conference on Artificial
  Intelligence}, volume~2, pages 1181--1188, 2008.

\bibitem{levie1982effects}
W.~H. Levie and R.~Lentz.
\newblock Effects of text illustrations: A review of research.
\newblock {\em ECTJ}, 30(4):195--232, 1982.

\bibitem{lin2013comparative}
J.~Lin, M.~Benisch, N.~Sadeh, J.~Niu, J.~Hong, B.~Lu, and S.~Guo.
\newblock A comparative study of location-sharing privacy preferences in the
  united states and china.
\newblock {\em Personal and Ubiquitous Computing}, 17(4):697--711, 2013.

\bibitem{lowe2004interrogation}
R.~Lowe.
\newblock Interrogation of a dynamic visualization during learning.
\newblock {\em Learning and Instruction}, 14(3):257--274, 2004.

\bibitem{luce1964simultaneous}
R.~D. Luce and J.~W. Tukey.
\newblock Simultaneous conjoint measurement: A new type of fundamental
  measurement.
\newblock {\em Journal of Mathematical Psychology}, 1(1):1--27, 1964.

\bibitem{mayer2001learning}
R.~E. Mayer and P.~Chandler.
\newblock When learning is just a click away: Does simple user interaction
  foster deeper understanding of multimedia messages?
\newblock {\em Journal of Educational Psychology}, 93(2):390--397, 2001.

\bibitem{mayer1990illustration}
R.~E. Mayer and J.~K. Gallini.
\newblock When is an illustration worth ten thousand words?
\newblock {\em Journal of Educational Psychology}, 82(4):715--726, 1990.

\bibitem{morrison2001effectiveness}
J.~B. Morrison and B.~Tversky.
\newblock The (in) effectiveness of animation in instruction.
\newblock In {\em CHI'01 Extended Abstracts on Human Factors in Computing
  Systems}, pages 377--378, 2001.

\bibitem{oberski2020differential}
D.~L. Oberski and F.~Kreuter.
\newblock Differential privacy and social science: An urgent puzzle.
\newblock {\em Harvard Data Science Review}, 2(1), 2020.

\bibitem{paivio2006dual}
A.~Paivio, J.~Clark, et~al.
\newblock Dual coding theory and education.
\newblock In {\em Draft chapter presented at the conference on Pathways to
  Literacy Achievement for High Poverty Children at The University of Michigan
  School of Education}. Citeseer, 2006.

\bibitem{roth2019honeycrisp}
E.~Roth, D.~Noble, B.~H. Falk, and A.~Haeberlen.
\newblock Honeycrisp: large-scale differentially private aggregation without a
  trusted core.
\newblock In {\em Proceedings of the 27th ACM Symposium on Operating Systems
  Principles}, pages 196--210, 2019.

\bibitem{roth2020orchard}
E.~Roth, H.~Zhang, A.~Haeberlen, and B.~C. Pierce.
\newblock Orchard: Differentially private analytics at scale.
\newblock In {\em 14th $\{$USENIX$\}$ Symposium on Operating Systems Design and
  Implementation ($\{$OSDI$\}$ 20)}, pages 1065--1081, 2020.

\bibitem{chowdhury2019outis}
A.~Roy~Chowdhury, C.~Wang, X.~He, A.~Machanavajjhala, and S.~Jha.
\newblock Crypt$\epsilon$: Crypto-assisted differential privacy on untrusted
  servers.
\newblock In {\em Proceedings of the 2020 ACM SIGMOD International Conference
  on Management of Data}, pages 603--619, 2020.

\bibitem{song2015privacyguard}
Y.~Song and U.~Hengartner.
\newblock Privacyguard: A vpn-based platform to detect information leakage on
  android devices.
\newblock In {\em Proceedings of the 5th Annual ACM CCS Workshop on Security
  and Privacy in Smartphones and Mobile Devices}, pages 15--26, 2015.

\bibitem{to2014framework}
H.~To, G.~Ghinita, and C.~Shahabi.
\newblock A framework for protecting worker location privacy in spatial
  crowdsourcing.
\newblock {\em Proceedings of the VLDB Endowment}, 7(10):919--930, 2014.

\bibitem{tversky2001spatial}
B.~Tversky.
\newblock Spatial schemas in depictions.
\newblock In {\em Spatial schemas and abstract thought}, volume~79, page 111,
  2001.

\bibitem{nylocation2018}
J.~Valentino-DeVries, N.~Singer, M.~H. Keller, and A.~Krolik.
\newblock Your apps know where you were last night, and they’re not keeping
  it secret.
\newblock {\em New York Times}, 10:2018, 2018.

\bibitem{Warner65}
S.~L. Warner.
\newblock Randomized response: A survey technique for eliminating evasive
  answer bias.
\newblock {\em Journal of the American Statistical Association},
  60(309):63--69, 1965.

\bibitem{warner1965randomized}
S.~L. Warner.
\newblock Randomized response: A survey technique for eliminating evasive
  answer bias.
\newblock {\em Journal of the American Statistical Association},
  60(309):63--69, 1965.

\bibitem{xiong2019arxiv}
A.~Xiong, T.~Wang, N.~Li, and S.~Jha.
\newblock Towards effective differential privacy communication for users’
  data sharing decision and comprehension.
\newblock In {\em 2020 IEEE Symposium on Security and Privacy (SP)}, pages
  392--410. IEEE, 2020.

\bibitem{zhu2017differentially}
T.~Zhu, G.~Li, W.~Zhou, and S.~Y. Philip.
\newblock Differentially private data publishing and analysis: A survey.
\newblock {\em IEEE Transactions on Knowledge and Data Engineering},
  29(8):1619--1638, 2017.

\end{thebibliography}
}

\appendix
\section{APPENDIX A: Model Illustrations} 
\label{sec:app_a}
{\small \subsection{Illustrations in Experiment 1}\label{sec:illstruation_i}}

\subsubsection{Central DP}
{\small \textbf{\\Static Illustration:}}

\begin{figure}[H]
     \centering
     \begin{subfigure}{0.41\textwidth}
         \includegraphics[width=\textwidth]{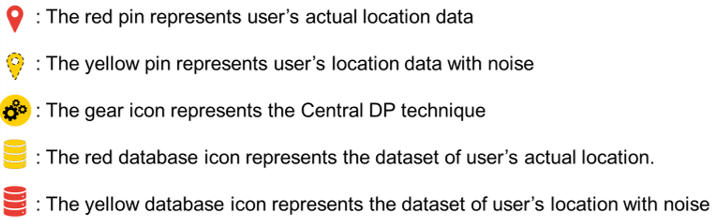} 
         \caption{\small Legend}
     \end{subfigure}
\end{figure}

\begin{figure}[H]
    \ContinuedFloat
    \begin{subfigure}{0.45\textwidth}
         \includegraphics[width=\textwidth]{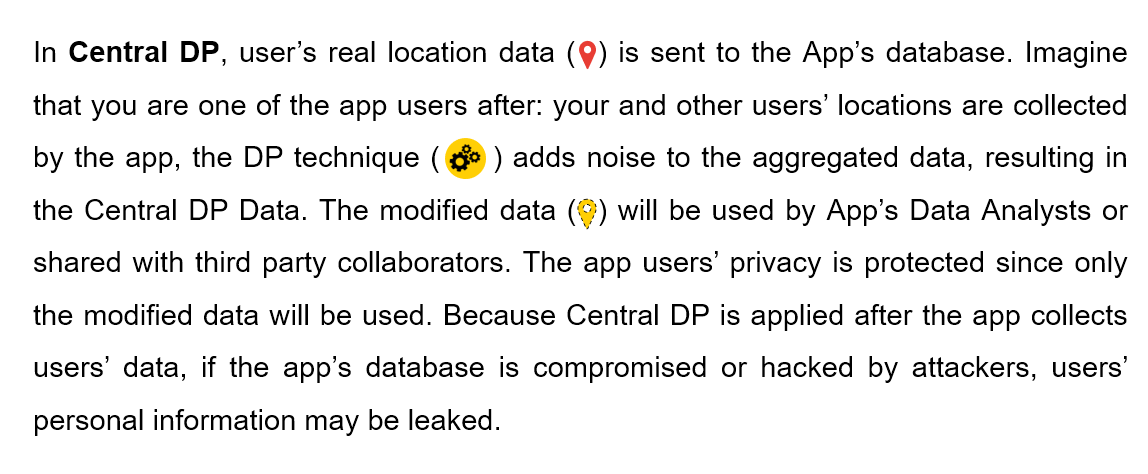} 
         \caption{\small Text description}
     \end{subfigure}
\end{figure}

\begin{figure}[H]
    \ContinuedFloat
    \begin{subfigure}{0.45\textwidth}
         \includegraphics[width=\textwidth]{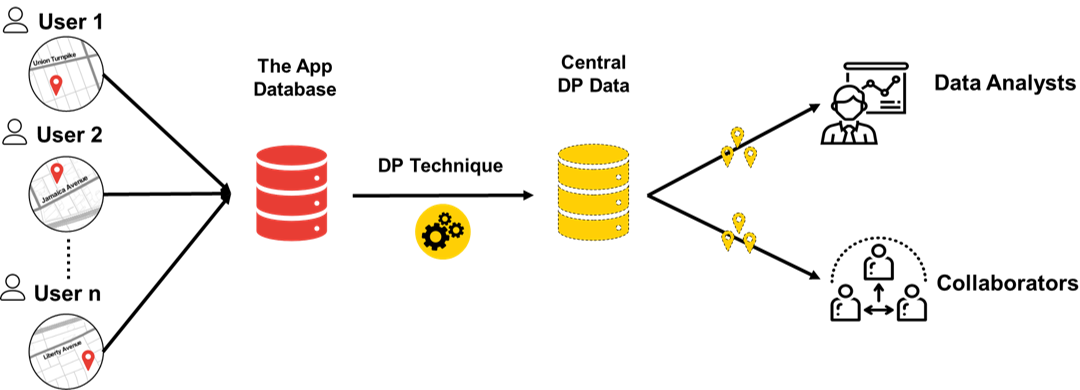}
         \caption{\small Data flow}
     \end{subfigure}
\end{figure}

\begin{figure}[H]
    \ContinuedFloat
    \begin{subfigure}{0.45\textwidth}
        \parbox{\textwidth}{\small This is a comparison between visualization produced with actual data and Central DP data. A small random amount of noise blurs out the three records in the circle without affecting the large scale patterns. Central DP keeps everyone in the dataset from being individually identifiable while maintains the usefulness of the dataset.}
         \includegraphics[width=\textwidth]{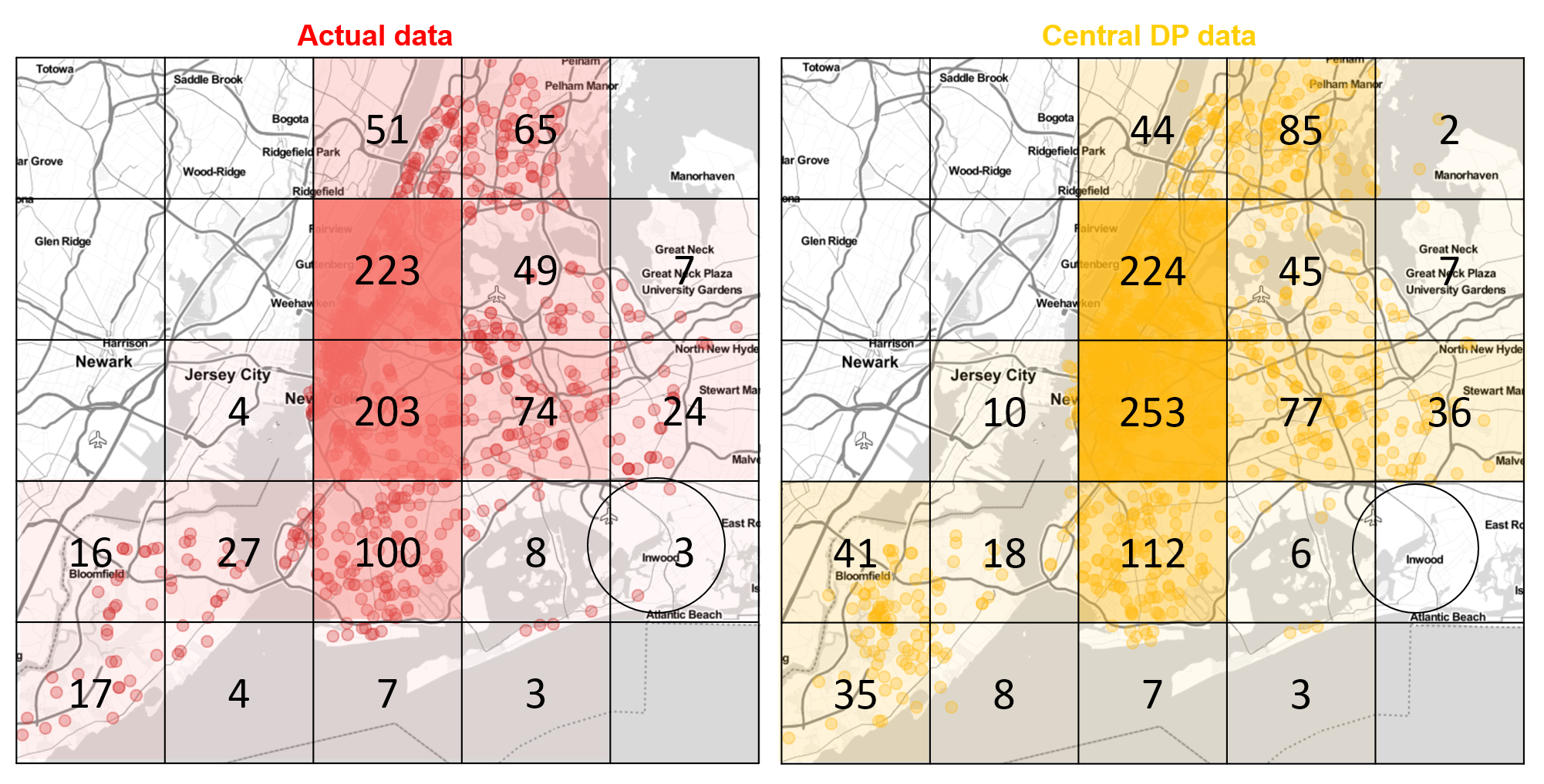}
         \caption{Utility}
     \end{subfigure}
        \caption{\small Illustrations of the Central DP model evaluated in the Experiment 1.}
        \label{fig:central_exp}
\end{figure}

\textbf{\small Animation Link:}

{\small https://youtu.be/2NvoryqUli8 }

\vspace{1.5mm}

\subsubsection{Local DP}
\vspace{2mm}
{\small\textbf{\\Static Illustration:}}
\begin{figure}[H]
     \centering
     \begin{subfigure}{0.45\textwidth}
         \includegraphics[width=\textwidth]{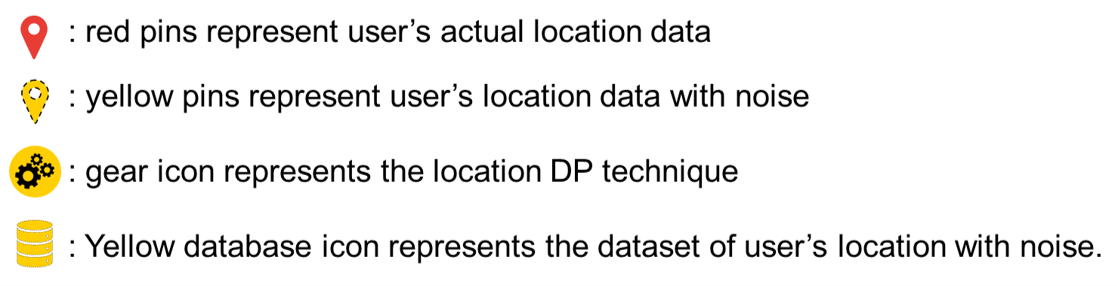} 
         \caption{\small Legend}
     \end{subfigure}
\end{figure}

\begin{figure}[H]
    \ContinuedFloat
    \begin{subfigure}{0.45\textwidth}
         \includegraphics[width=\textwidth]{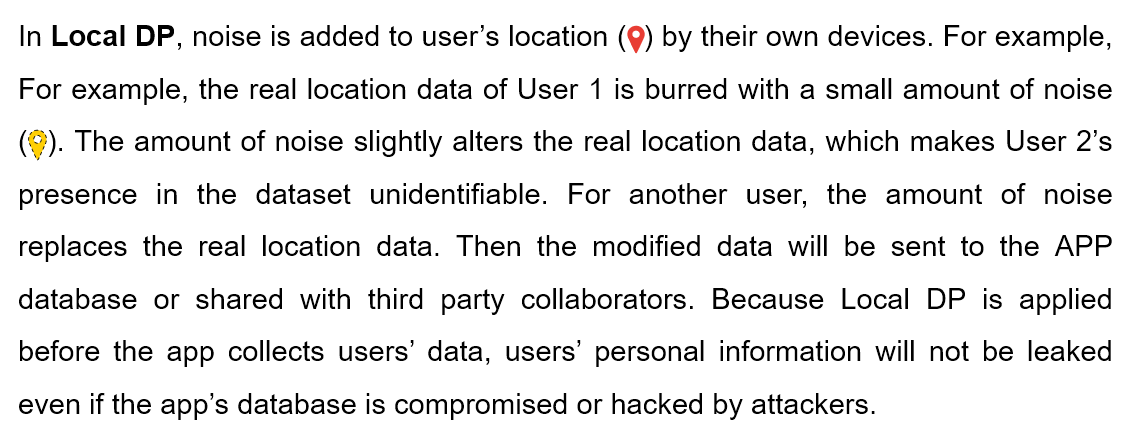} 
         \caption{\small Text description}
     \end{subfigure}
\end{figure}

\begin{figure}[H]
    \ContinuedFloat
    \begin{subfigure}{0.45\textwidth}
         \includegraphics[width=\textwidth]{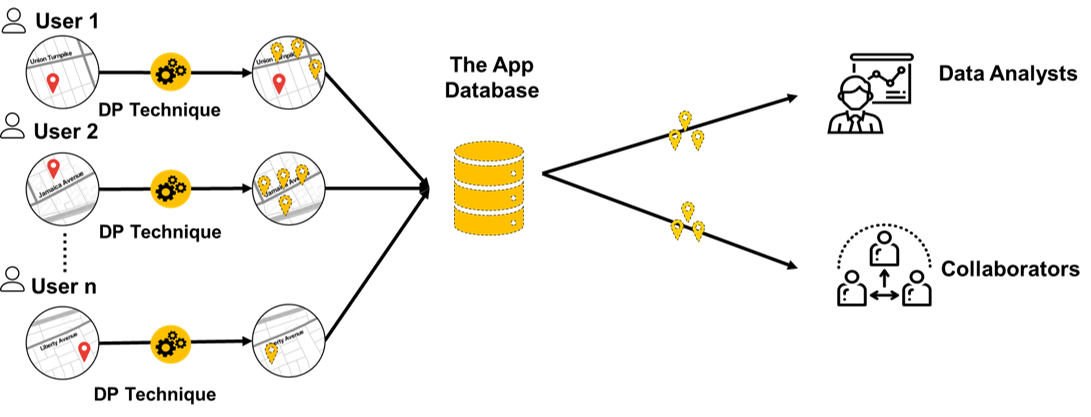}
         \caption{\small Data flow}
     \end{subfigure}
\end{figure}

\begin{figure}[H]
    \ContinuedFloat
    \begin{subfigure}{0.45\textwidth}
        \parbox{\textwidth}{\small This is a comparison between visualization produced with actual data and Local DP data. Local DP keeps everyone in the dataset from being individually identifiable, while maintains the usefulness of the dataset. The degree of noise added affects both how well individual data is protected and how useful the data set is. Different amounts of noise from each user provides strong privacy protection, but the large-scale pattern is somewhat impacted.}
         \includegraphics[width=\textwidth]{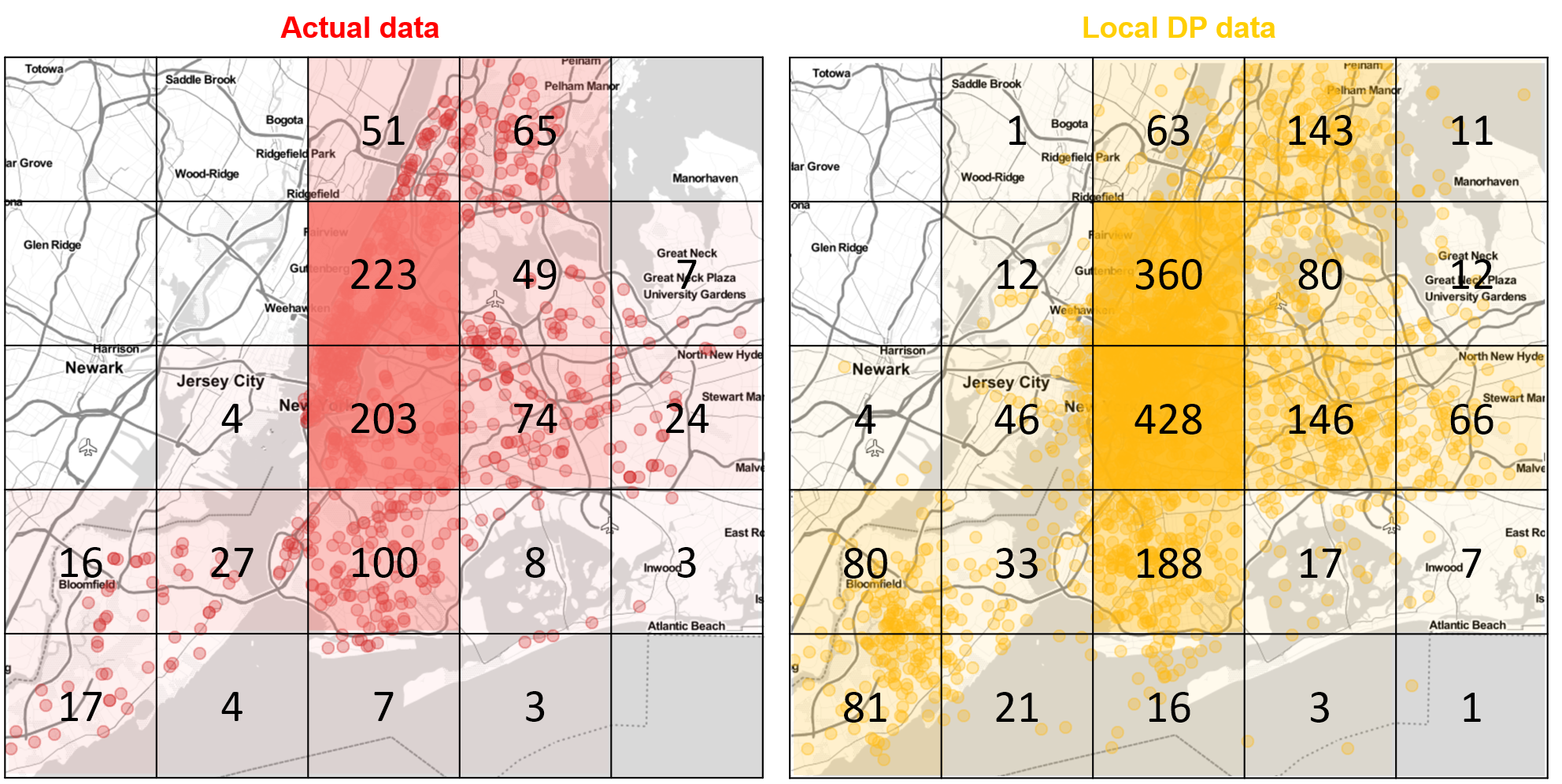}
         \caption{Utility}
     \end{subfigure}
        \caption{\small Illustrations of the Local DP model evaluated in the Experiment 1.}
        \label{fig:local_exp}
\end{figure}

\textbf{\small Animation Link:}

{\small https://youtu.be/DEYv9QkYWF0}

\vspace{1.5mm}

\subsubsection{\small Shuffler DP}
{\small \textbf{\\Static Illustration:}}

\begin{figure}[H]
     \centering
     \begin{subfigure}{0.45\textwidth}
         \includegraphics[width=\textwidth]{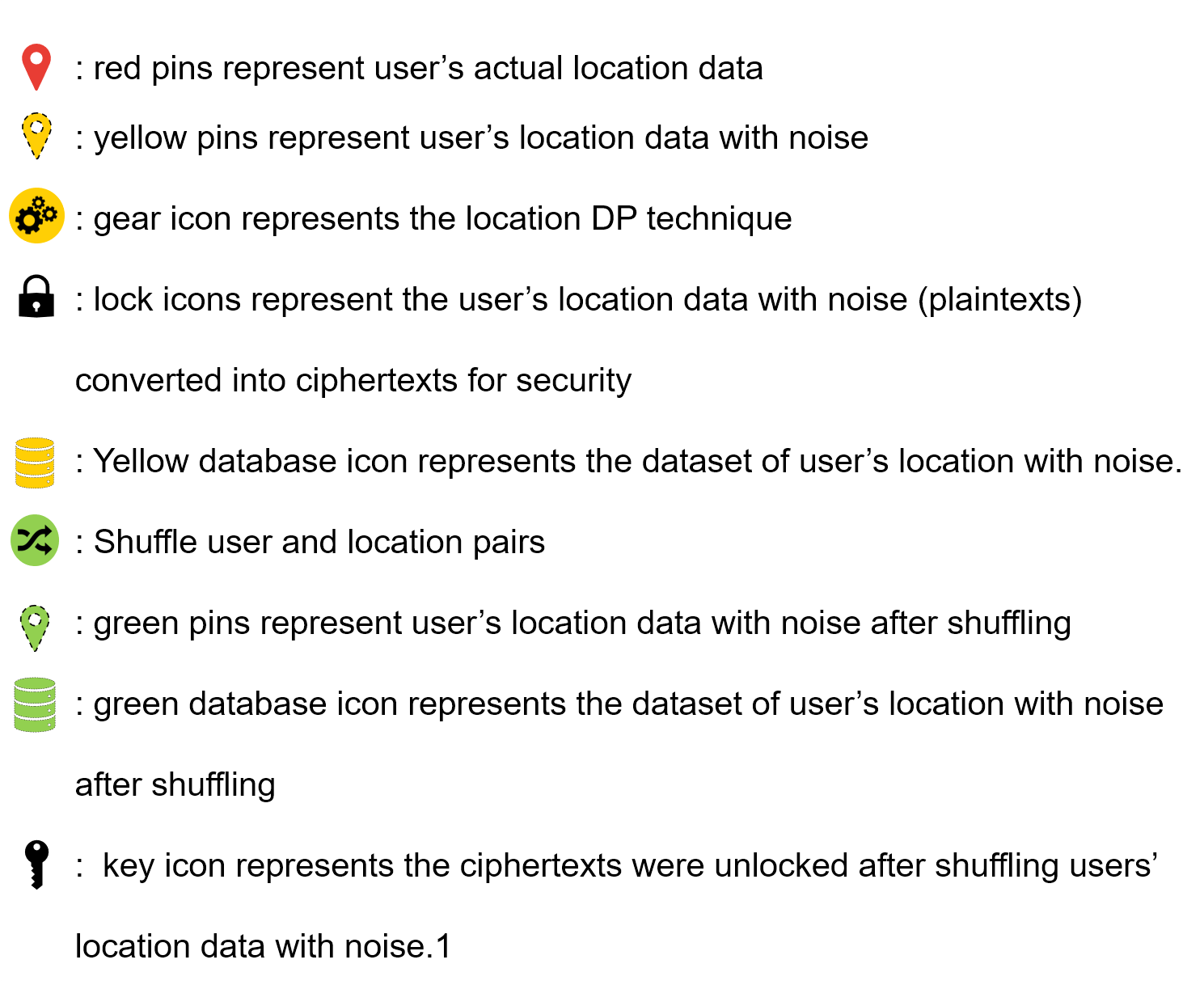} 
         \caption{\small Legend}
     \end{subfigure}
\end{figure}

\begin{figure}[H]
    \ContinuedFloat
    \begin{subfigure}{0.48\textwidth}
         \includegraphics[width=\textwidth]{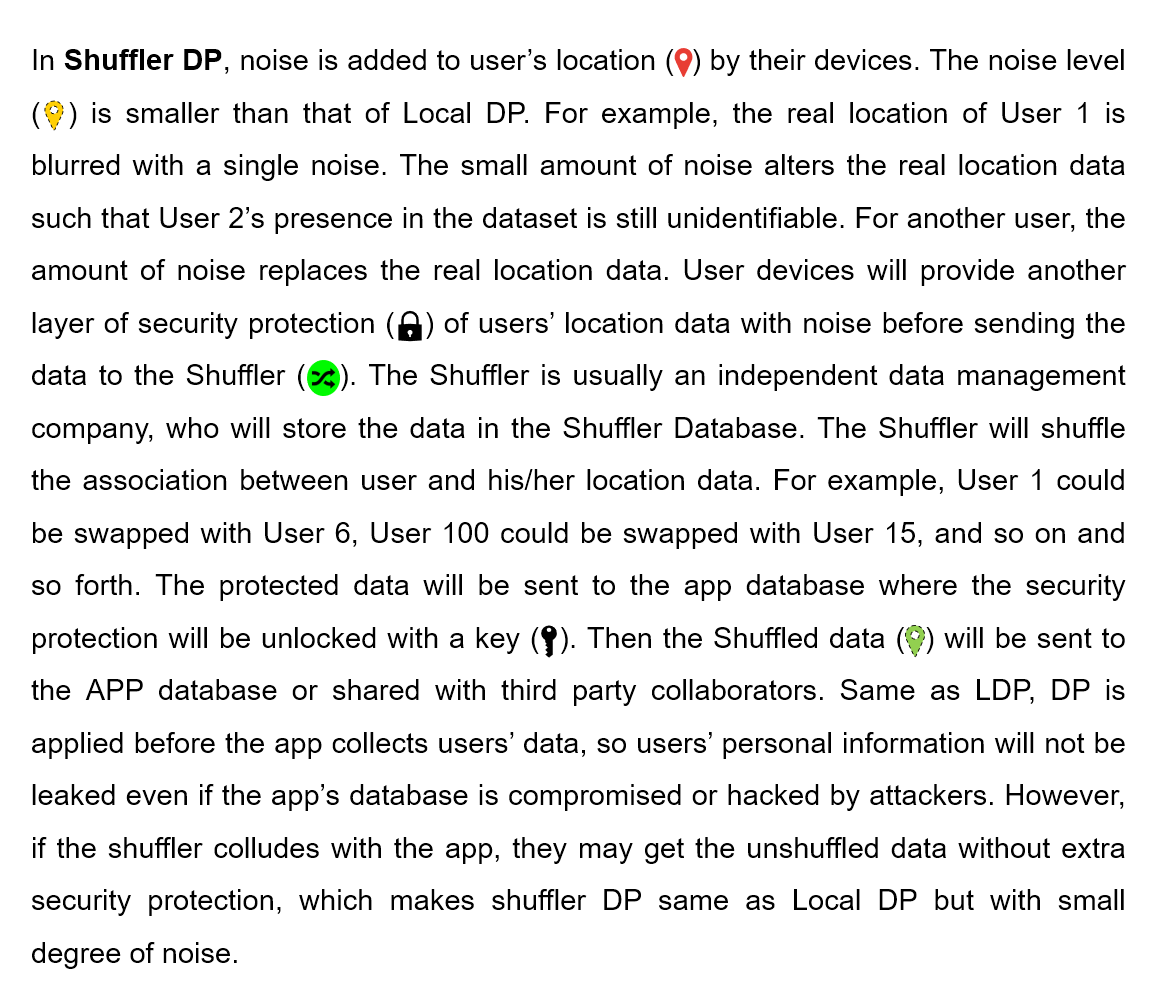} 
         \caption{\small Text description}
     \end{subfigure}
\end{figure}

\begin{figure}[H]
    \ContinuedFloat
    \begin{subfigure}{0.45\textwidth}
         \includegraphics[width=\textwidth]{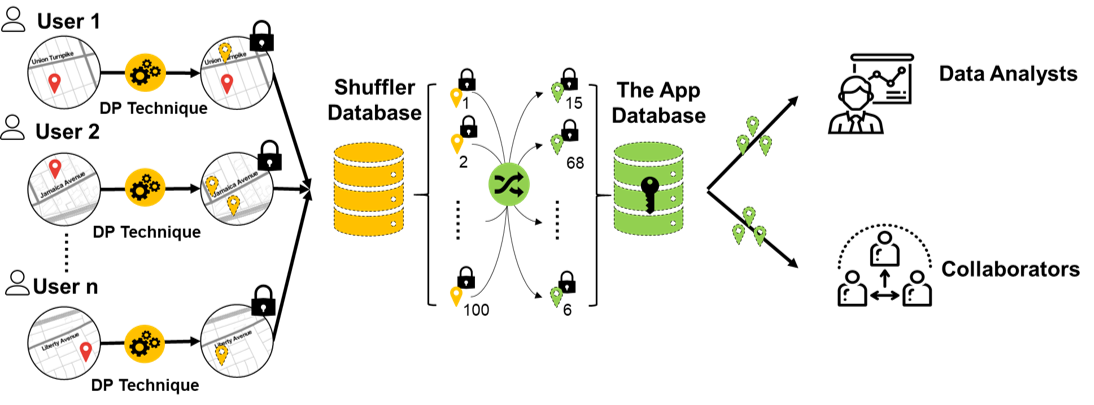}
         \caption{\small Data flow}
     \end{subfigure}
\end{figure}

\begin{figure}[H]
    \ContinuedFloat
    \begin{subfigure}{0.45\textwidth}
        \parbox{\textwidth}{\small This is a comparison between visualization produced with actual data and Shuffler DP data. Smaller amount of noise makes Shuffler DP maintains the large-scale pattern more similar to actual data than Local DP.}
         \includegraphics[width=\textwidth]{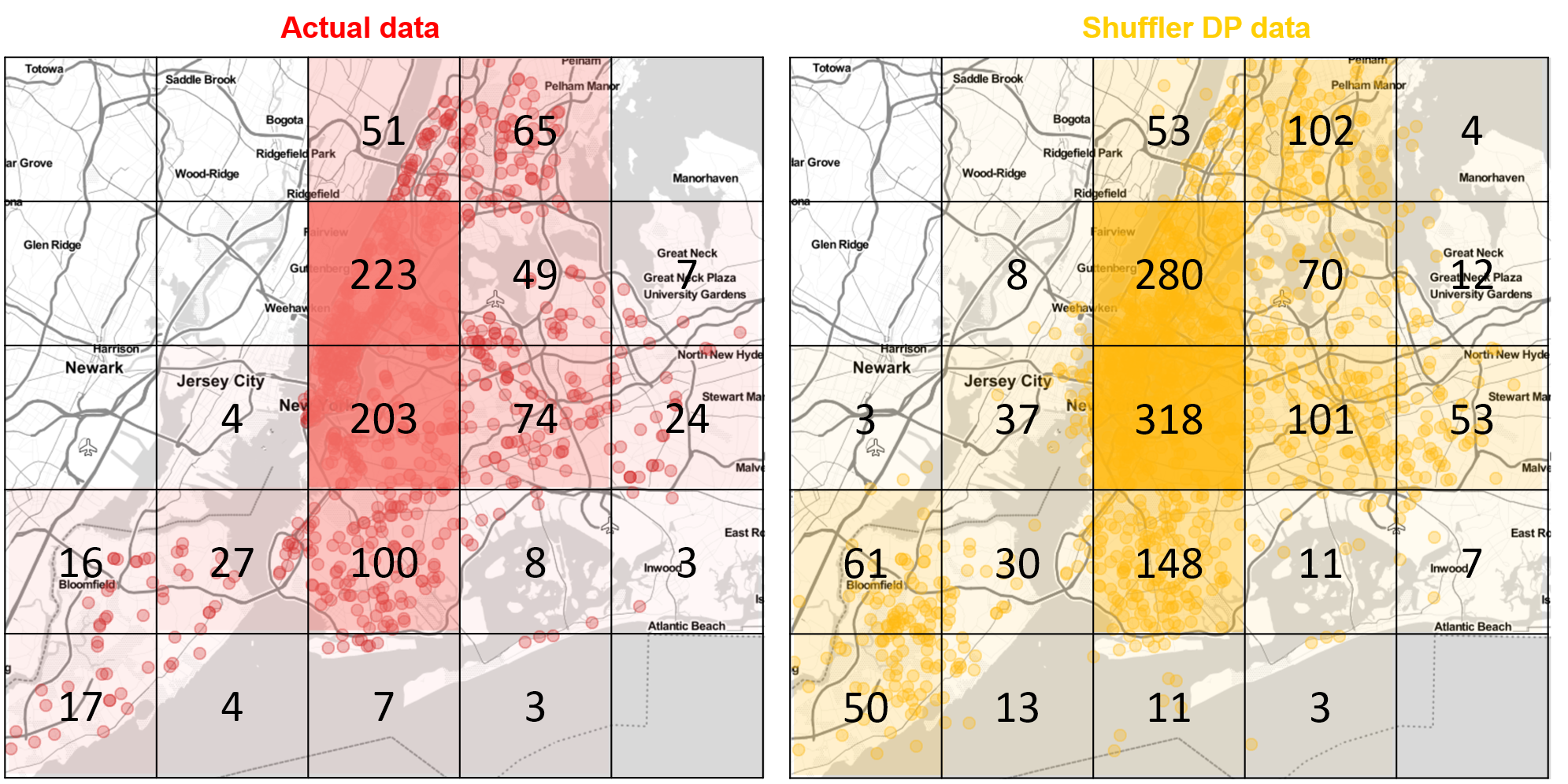}
         \caption{Utility}
     \end{subfigure}
     \vspace{-0.1cm}
        \caption{\small Illustrations of the Shuffler DP model evaluated in Experiment 1.}
        \label{fig:shuffler_exp}
\end{figure}

\textbf{\small Animation Link:}

{\small https://youtu.be/7wBxGsvpZsU}

\subsection{\small Illustration of Experiment 2}\label{sec:illstruation_ii}
\vspace{-30mm}

\begin{figure}[H]
\includegraphics[width=0.48\textwidth]{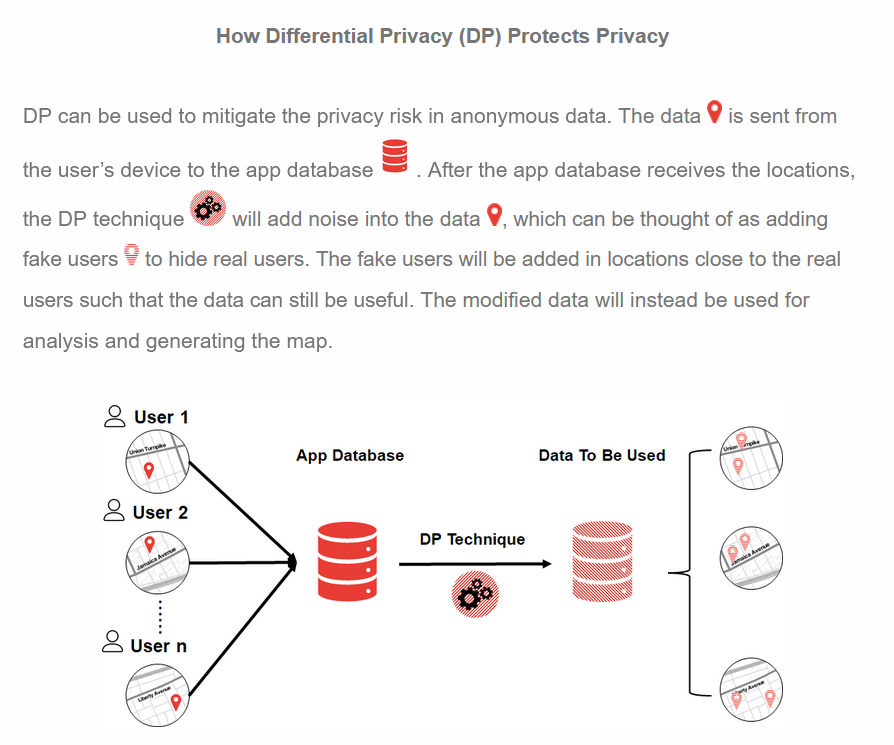}
\caption{\small Illustration of the Central DP model evaluated in the Experiment 2 }
\label{fig:central_scenario}
\end{figure}


\begin{figure}[H]
\includegraphics[width=0.48\textwidth]{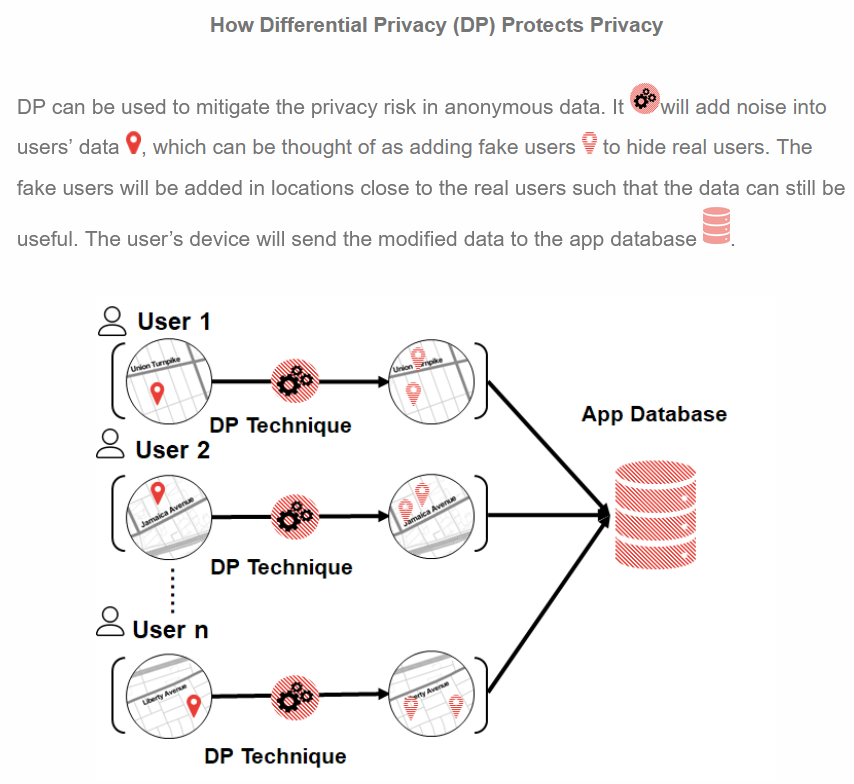}
\caption{\small Illustration of the Local DP model evaluated in the Experiment 2 }
\label{fig:local_scenario}
\end{figure}


\begin{figure}[H]
     \centering
     \begin{subfigure}{0.45\textwidth}
         \includegraphics[width=\textwidth]{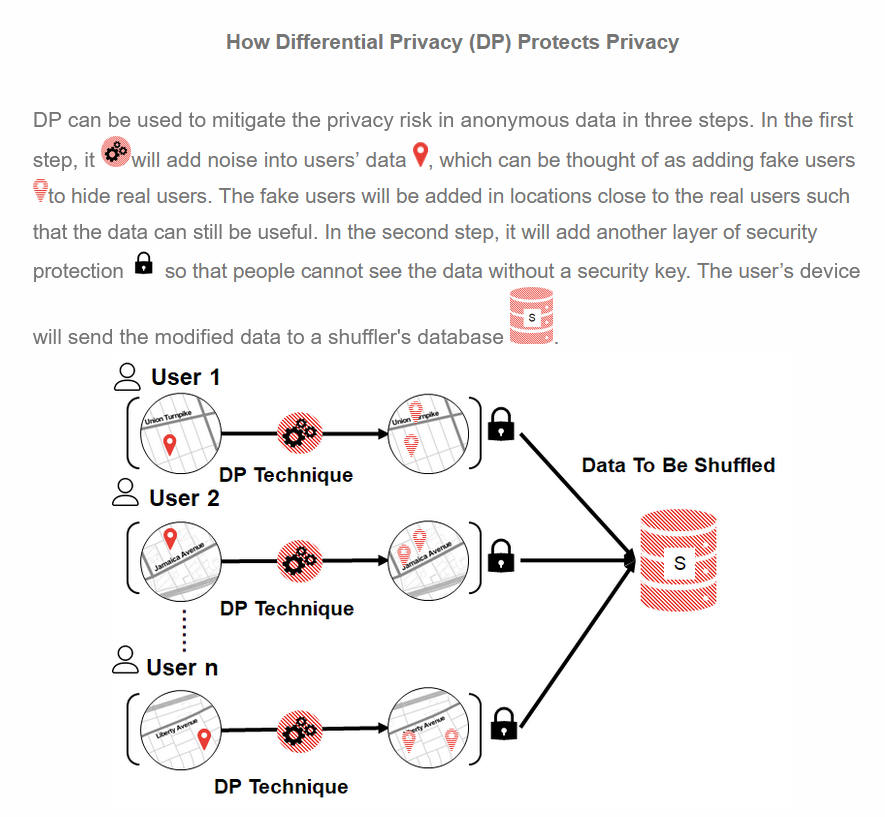} 
     \end{subfigure}
\end{figure}

\begin{figure}[H]
    \ContinuedFloat
     \centering
     \begin{subfigure}{0.45\textwidth}
         \includegraphics[width=\textwidth]{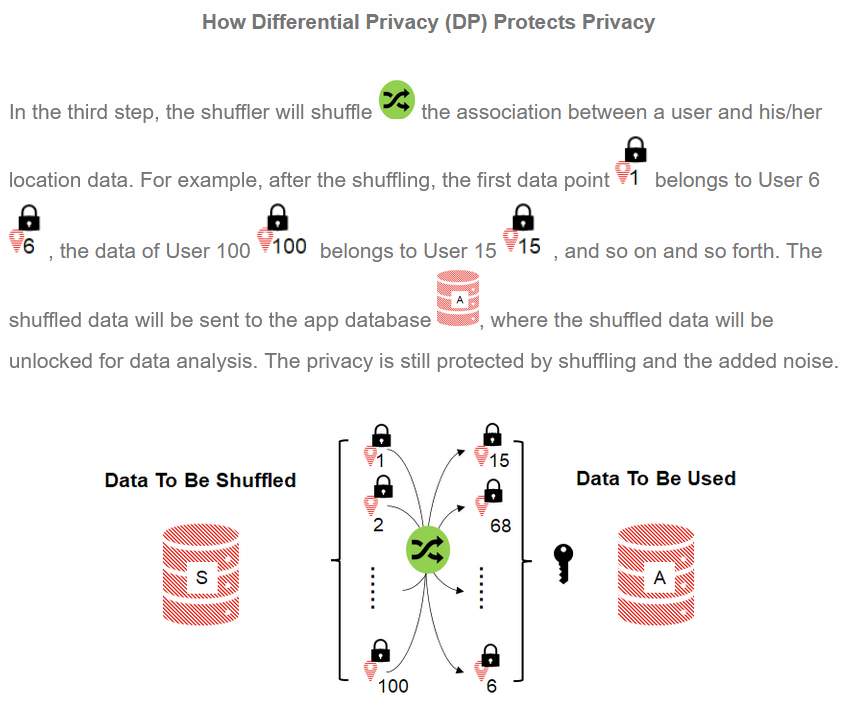}
     \end{subfigure}
     \vspace{-0.3cm}
        \caption{\small Illustrations of the Shuffler DP model evaluated in Experiment 2.}
        \label{fig:shuffler_scenario}
\end{figure}

\section{APPENDIX B: Survey Protocol} \label{sec:app_b}

\subsection{Survey Protocol of Experiment 1}
\label{sec:survey_protocol}
{\small \noindent\textbf{Central DP:}}
\par\noindent\rule{0.48\textwidth}{0.2pt}

{\small \noindent\textbf{[Illustration/animation of Central DP - first presentation]}  }  
\vspace{2mm}

{\small \noindent\textbf{[Comprehension]}}

{\small \question{CQ1-1} Suppose that you have your location information collected by the app, but your location information was collected using Central DP. If an attacker gets access to the database of the app, will the attacker be able to see your real location information?}

{\small \begin{itemize} [noitemsep,nosep]
    \item[$\circ$] \textbf{Yes}
    \item[$\circ$] Unsure
    \item[$\circ$] No
    \item[$\circ$] Prefer not to answer
\end{itemize}
\noindent\textbf{[Comprehension feedback]}}

{\small \textbf{Your answer is correct (incorrect).} Since understanding the DP technique is critical for answering questions afterward, please view the illustration again on the next page.}

{\small \noindent\textbf{[Illustration/animation of Central DP - second presentation]} }
    
    
{\small \noindent\textbf{[Privacy Protection and Utility Perception]}}
    
    {\small Please indicate your disagreement or agreement with the following statements on a 7-point Likert Scale, ``1'" means ``Strongly disagree", and ``7" means ``Strongly agree". }
    
{\small \question{PQ1} Using \textbf{Central DP}, the collected user location data is still very useful. }
{\small \begin{itemize} [noitemsep,nosep]
    \item[$\circ$] Strongly disagree (1)
    \item[$\circ$] Disagree (2)
    \item[$\circ$] More or less disagree (3)
    \item[$\circ$] Neither disagree or agree (4)
    \item[$\circ$] More or less agree (5)
    \item[$\circ$] Agree (6)
    \item[$\circ$] Strongly agree (7)
\end{itemize}}
    
{\small \question{UQ1} Using \textbf{Central DP}, the collected user location data achieves desirable privacy preservation and security protection.  }
{\small 
\begin{itemize} [noitemsep,nosep]
    \item[$\circ$] Strongly disagree (1)
    \item[$\circ$] Disagree (2)
    \item[$\circ$] More or less disagree (3)
    \item[$\circ$] Neither disagree or agree (4)
    \item[$\circ$] More or less agree (5)
    \item[$\circ$] Agree (6)
    \item[$\circ$] Strongly agree (7)
\end{itemize}}


{\small \noindent\textbf{[Data-Sharing Decision]}}
    
    {\small Please indicate your disagreement or agreement with the following statements on a 7-point Likert Scale, ``1'" means ``Strongly disagree", and ``7" means ``Strongly agree". }
    
{\small \question{SQ1-1} If the app collects information through \textbf{Central DP} and uses it for companies to make relevant recommendations, I am willing to share my information.}

{\small \begin{itemize} [noitemsep,nosep]
    \item[$\circ$] Strongly disagree (1)
    \item[$\circ$] Disagree (2)
    \item[$\circ$] More or less disagree (3)
    \item[$\circ$] Neither disagree or agree (4)
    \item[$\circ$] More or less agree (5)
    \item[$\circ$] Agree (6)
    \item[$\circ$] Strongly agree (7)
\end{itemize}}
    
{\small \question{SQ1-2} If the app collects information through \textbf{Central DP} and uses it for research in disease control and prevention, I am willing to share my information. 
\begin{itemize} [noitemsep,nosep]
    \item[$\circ$] Strongly disagree (1)
    \item[$\circ$] Disagree (2)
    \item[$\circ$] More or less disagree (3)
    \item[$\circ$] Neither disagree or agree (4)
    \item[$\circ$] More or less agree (5)
    \item[$\circ$] Agree (6)
    \item[$\circ$] Strongly agree (7)
\end{itemize}}
    
{\small \noindent\textbf{Local DP:}}
\par\noindent\rule{0.48\textwidth}{0.2pt}

{\small \noindent\textbf{[Illustration/animation of Local DP - first presentation]} }   

{\small \noindent\textbf{[Comprehension]}}

{\small \question{CQ1-2} Suppose that you have your location information collected by the app, but your location information was collected using Local DP. For the third party with which the app shared data, will they be able to see the real answer that you submitted?}
{\small \begin{itemize} [noitemsep,nosep]
    \item[$\circ$] Yes 
    \item[$\circ$] Unsure
    \item[$\circ$] \textbf{No}
    \item[$\circ$] Prefer not to answer
\end{itemize}}
{\small \noindent\textbf{[Comprehension feedback]}}


{\small \noindent\textbf{[Illustration/animation of Local DP - second presentation]}}
    
    
{\small \noindent\textbf{[Privacy Protection and Utility Perception]}}


{\small \noindent\textbf{[Data-Sharing Decision]}}

{\small \noindent\textbf{Shuffler DP:}}
\par\noindent\rule{0.48\textwidth}{0.2pt}

{\small \noindent\textbf{[Illustration/animation of Shuffler DP - first presentation]} }   

{\small \noindent\textbf{[Comprehension]}}

{\small \question{CQ1-3} Suppose that you have your location information collected by the app, but your location information was collected using  Shuffler DP. For the third party with which the app shared data, will the data still be useful?}
{\small \begin{itemize} [noitemsep,nosep]
    \item[$\circ$] \textbf{Yes}
    \item[$\circ$] Unsure
    \item[$\circ$] No
    \item[$\circ$] Prefer not to answer
\end{itemize}}
{\small \noindent\textbf{[Comprehension feedback]}}


\noindent\textbf{[Illustration/animation of Shuffler DP - second presentation]}
    
    
{\small \noindent\textbf{[Privacy Protection and Utility Perception]}}


{\small \noindent\textbf{[Data-Sharing Decision]}}


\par\noindent\rule{0.48\textwidth}{0.2pt}

{\small \noindent\textbf{[Comprehension]}(Randomized)}

CQ1-1.

CQ1-2.

CQ1-3.

\vspace{2mm}

\subsection{Survey Protocol of Experiment 2}
\label{sec:survey_protocol_II}

\textbf{[Privacy Risks]}

\begin{figure}[H]
     \centering
     \begin{subfigure}{0.42\textwidth}
         \includegraphics[width=\textwidth]{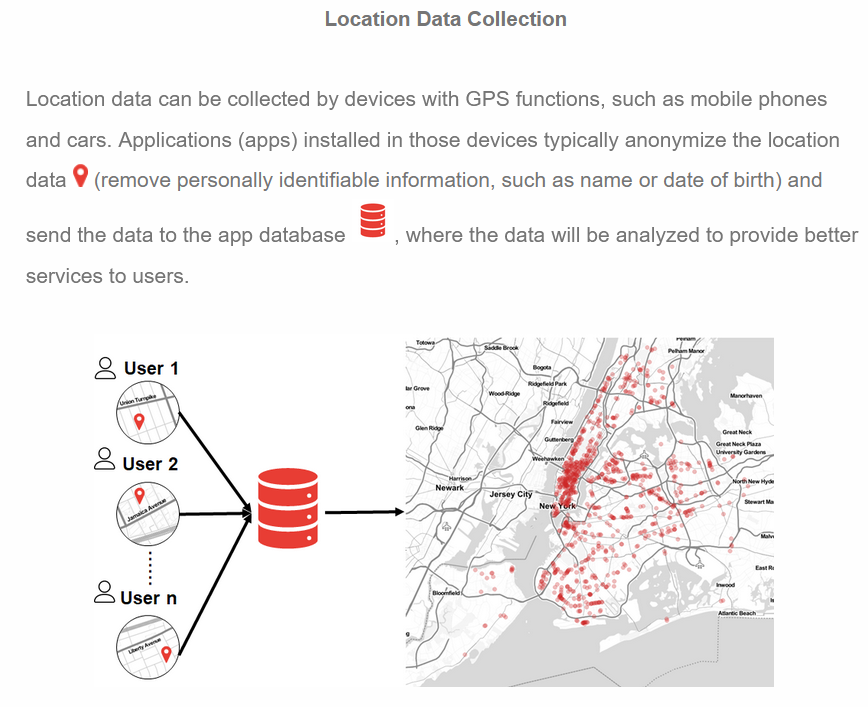} 
     \end{subfigure}
\end{figure}

\begin{figure}[H]
    \ContinuedFloat
    \begin{subfigure}{0.42\textwidth}
         \includegraphics[width=\textwidth]{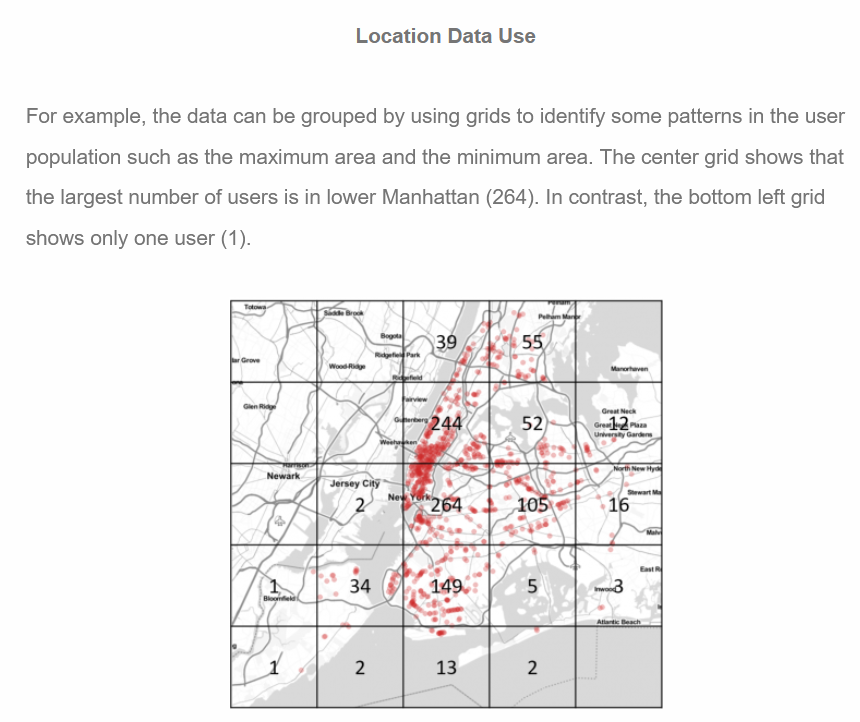}
     \end{subfigure}
\end{figure}

\begin{figure}[H]
    \ContinuedFloat
    \begin{subfigure}{0.42\textwidth}
         \includegraphics[width=\textwidth]{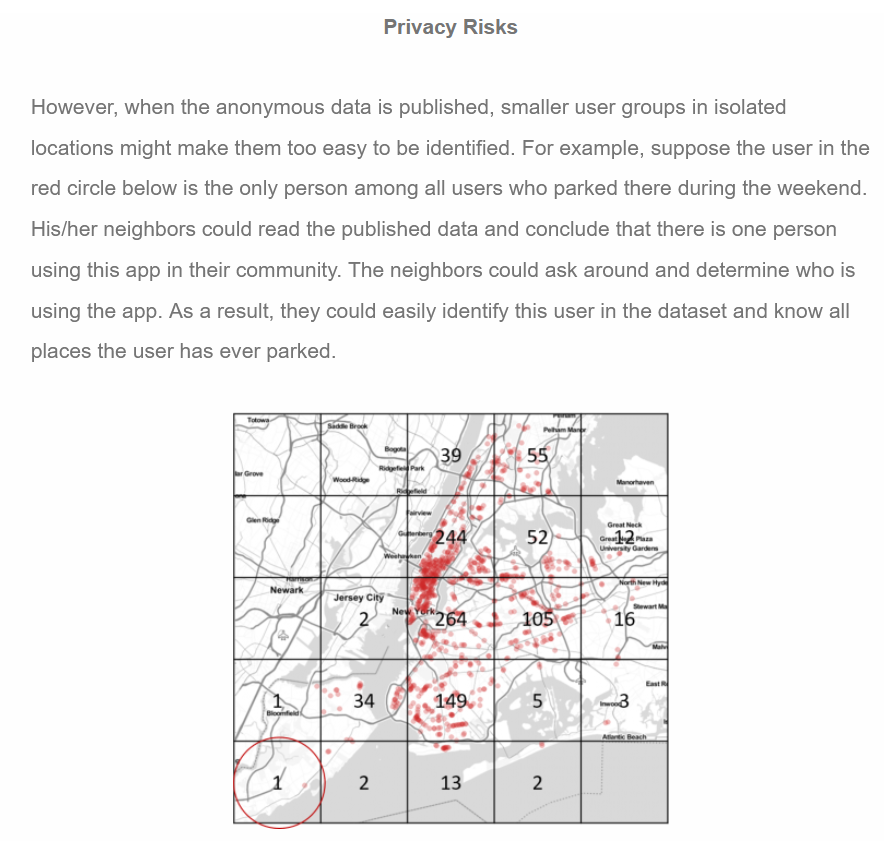}
     \end{subfigure}
     \vspace{-0.3cm}
        \caption{Illustration of the privacy risks.}
        \label{fig:privacy_risks}
\end{figure}

\textbf{[Comprehension]}
\question{CQ2-1} One app company wants to share its user location data with a third party. The app company fully anonymized the data (it would not contain personally identifiable information, such as name or date of birth). Will there be a risk for users in the anonymized dataset to be identified?
\begin{itemize} [noitemsep,nosep]
    \item[$\circ$] \textbf{Yes} 
    \item[$\circ$] Unsure
    \item[$\circ$] No
    \item[$\circ$] Prefer not to answer
\end{itemize}
\vspace{2mm}

{\small \textbf{[Comprehension feedback]}}
{\small \textbf{Your answer is correct (incorrect).} Anonymization alone cannot provide sufficient privacy protection.}

{\small \textbf{[Illustration of Differential Privacy - Central, Local, or Shuffler]}}

{\small \textbf{[Comprehension]}}
{\small \question{CQ2-2} An app company decided to deploy DP to improve the privacy protection of its users. With the deployment of DP, will the initial data received by the app company contain any noise?}
{\small \begin{itemize} [noitemsep,nosep]
    \item[$\circ$] \textbf{Yes} for Local/Shuffler
    \item[$\circ$] Unsure
    \item[$\circ$] \textbf{No} for Central
    \item[$\circ$] Prefer not to answer
\end{itemize}}

{\small \textbf{[Comprehension feedback]}}
{\small \textbf{Your answer is correct (incorrect).} Noise is added before the device sends the data.
}

{\small \question{CQ2-3} An app company decided to use DP to improve the privacy protection of its users. Which of the following would happen if the third step (shuffling) is omitted in the DP deployment? }
{\small \begin{itemize} [noitemsep,nosep]
    \item[$\circ$] The level of privacy protection will increase
    \item[$\circ$] Unsure
    \item[$\circ$] \textbf{The level of privacy protection will decrease}
    \item[$\circ$] Prefer not to answer
\end{itemize}}

{\small \textbf{[Illustrations of Noise Levels]}}

\begin{figure}[H]
     \centering
     \begin{subfigure}{0.3\textwidth}
         \includegraphics[width=\textwidth]{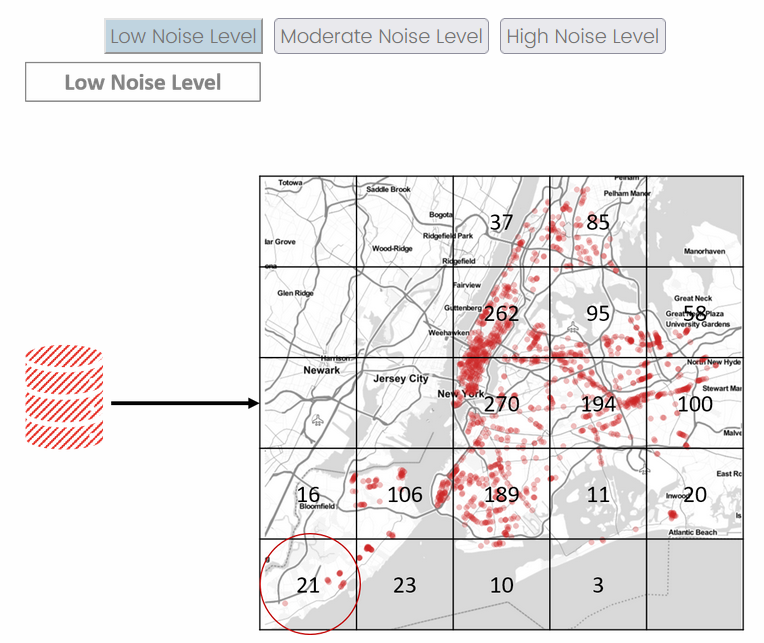} 
     \end{subfigure}
\end{figure}


\begin{figure}[H]
    \centering
    \ContinuedFloat
    \begin{subfigure}{0.3\textwidth}
         \includegraphics[width=\textwidth]{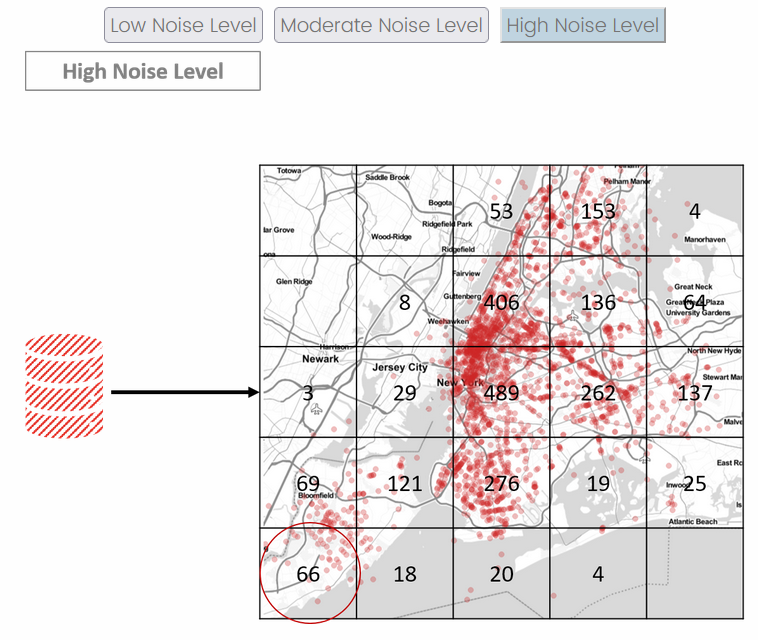}
     \end{subfigure}
     \vspace{-0.3cm}
        \caption{Illustrations of different levels of noise in DP.}
        \label{fig:noise_level}
\end{figure}

{\small \textbf{[Comprehension]}}
{\small \question{CQ2-4} An app company deployed DP to protect the privacy of its users. The app company now wants to provide more accurate location-based services to users. Which of the following option should the app company choose?}
{\small \begin{itemize} [noitemsep,nosep]
    \item[$\circ$] Increase the noise level in DP
    \item[$\circ$] Unsure
    \item[$\circ$] \textbf{Decrease the noise level in DP} 
    \item[$\circ$] Prefer not to answer
\end{itemize}}

{\small \textbf{[Comprehension feedback]}}
{\small \textbf{Your answer is correct (incorrect).} A lower noise level means that the data is more accurate.}

{\small \question{CQ2-5} An app company deployed DP to protect the privacy of its users. As a user of the app, you can set up the noise level that you prefer before using the app. If you prefer stronger privacy protection, which of the following option should you choose?}
{\small 
\begin{itemize} [noitemsep,nosep]
    \item[$\circ$] \textbf{Higher noise level} 
    \item[$\circ$] Unsure
    \item[$\circ$] Lower noise level
    \item[$\circ$] Prefer not to answer
\end{itemize}}

{\small \textbf{[Comprehension feedback]}}
{\small \textbf{Your answer is correct (incorrect).} A higher noise level means that the privacy is stronger.}

{\small \textbf{[Privacy and Accuracy Perception]}}

{\small Please indicate your disagreement or agreement with the following statements on a 7-point Likert Scale: ``1" means ``Strongly disagree", and ``7" means ``Strongly agree". }

{\small \question{UQ2} My expected accuracy of collected data can be reasonably maintained by the Differential Privacy (DP) technique.}

{\small 
\begin{itemize} [noitemsep,nosep]
    \item[$\circ$] Strongly disagree (1)
    \item[$\circ$] Disagree (2)
    \item[$\circ$] More or less disagree (3)
    \item[$\circ$] Neither disagree or agree (4)
    \item[$\circ$] More or less agree (5)
    \item[$\circ$] Agree (6)
    \item[$\circ$] Strongly agree (7)
\end{itemize}}

{\small \question{PQ2} My privacy of data collection and use can be reasonably achieved by the Differential Privacy (DP) technique.}

{\small \begin{itemize} [noitemsep,nosep]
    \item[$\circ$] Strongly disagree (1)
    \item[$\circ$] Disagree (2)
    \item[$\circ$] More or less disagree (3)
    \item[$\circ$] Neither disagree or agree (4)
    \item[$\circ$] More or less agree (5)
    \item[$\circ$] Agree (6)
    \item[$\circ$] Strongly agree (7)
\end{itemize}}

{\small \textbf{[Illustration of Differential Privacy and Noise Level - second presentation]}}

{\small \textbf{[Data-Disclosure Scenarios]}}

{\small \question{Electrical Vehicle Charging Scenario:} Suppose there is a dataset containing parking locations of electric vehicles (EVs). Using those location data, the city planners can make better decisions on where to build new charging stations for EVs. However, without extra protection from privacy-enhancing technologies, such as Differential Privacy (DP), the identities of those EV owners can be inferred from unique parking patterns of home locations and work places. Consequently, all the places where each EV owner has ever been can be revealed.}

{\small Imagine that you are one of those EV owners. You are asked to choose a noise level of the DP technique to protect your privacy while maintaining the accuracy of the whole dataset. You can use the three buttons on the right to view how different levels of noise will impact the accuracy. We also present the real data without noise on the left for your reference.}

\begin{figure}[H]
\includegraphics[width=0.45\textwidth]{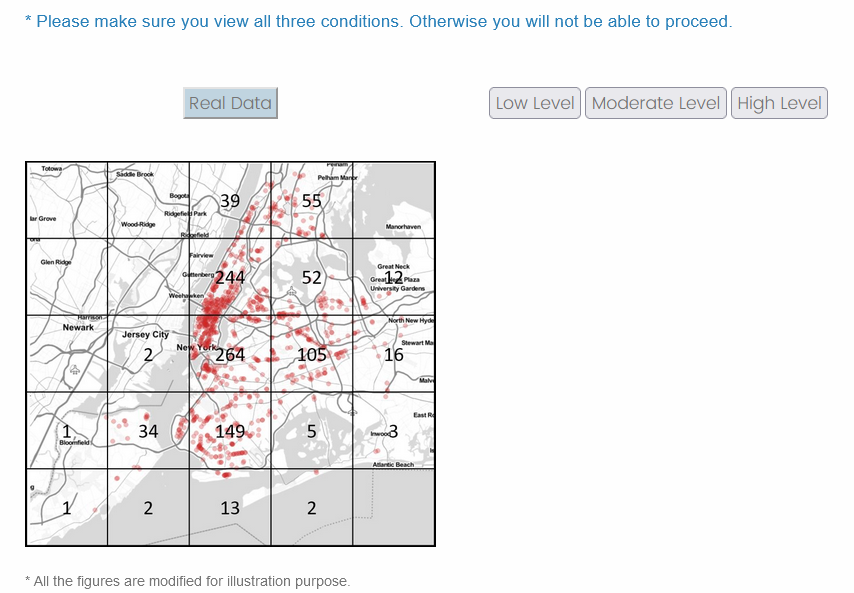}

{\small \caption{ Illustration of Different level of noise in the Scenario.}}
\label{fig:central_scenario}

\end{figure}

{\small \question{Disabled Parking Spaces Scenario:} Suppose there is a dataset including locations of disabled parking. Using those location data, the city planners can make better decisions on whether the current percentage of disabled parking spaces in different parking lots are appropriate. However, without extra protection from privacy-enhancing technologies, such as differential privacy (DP), the identities of those vehicle owners can be inferred from unique parking patterns of home locations and work places. Consequently, all the places where each vehicle owner has ever been can be revealed.}

{\small Imagine that you are one of those vehicle owners. You are asked to choose a noise level of the DP technique to protect your privacy while maintaining the accuracy of the whole dataset. You can use the three buttons on the right to view how different levels of noise will impact the accuracy. We also present the real data without noise on the left for your reference.} 

{\small \question{Company Advertisements Scenario:} Suppose there is a dataset including locations of vehicles. Using those location data, companies providing car-related services can decide better locations to post their advertisement. However, without extra protection from privacy-enhancing technologies, such as Differential Privacy (DP), the identities of those vehicle owners can be inferred from unique parking patterns of home locations and work places. Consequently, all the places where each vehicle owner has ever been can be revealed.}

{\small Imagine that you are one of those vehicle owners. You are asked to choose a noise level of the DP technique to protect your privacy while maintaining the accuracy of the whole dataset. You can use the three buttons on the right to view how different levels of noise will impact the accuracy. We also present the real data without noise on the left for your reference. }

{\small \question{Parking Garage Investment Scenario:}} 
{\small Suppose there is a dataset including parking locations of vehicles. Using those location data, investors can decide better locations to invest for new parking garages. However, without extra protection from privacy-enhancing technologies, such as Differential Privacy (DP), the identities of those vehicle owners can be inferred from unique parking patterns of home locations and work places. Consequently, all the places where each vehicle owner has ever been can be revealed.}

{\small Imagine that you are one of those vehicle owners. You are asked to choose a noise level of the DP technique to protect your privacy while maintaining the accuracy of the whole dataset. You can use the three buttons on the right to view how different levels of noise will impact the accuracy. We also present the real data without noise on the left for your reference.} 

{\small \question{SQ2\_1} Among the following options, which one would you prefer to be deployed for this location-based service?.}

{\small \begin{itemize} [noitemsep,nosep]
    \item[$\circ$] Low noise level 
    \item[$\circ$] Unsure
    \item[$\circ$] High noise level
    \item[$\circ$] Prefer not to answer
\end{itemize}}

{\small \question{SQ2\_2} Could you briefly explain why you chose \_?}


{\small \textbf{[Demographics]}}
\section{APPENDIX C: Pilot Study} \label{sec:app_c}

\end{document}